\documentclass[iop]{emulateapj}

\usepackage{epsfig}
\usepackage{graphicx}
\usepackage{epstopdf}
\usepackage{amsmath}
\usepackage{natbib}

\begin{document}

\title{Bayesian Inference of Polarized CMB Power Spectra from Interferometric Data}
\author{Ata Karakci$^{1}$,
 P.~M.~Sutter$^{2,3,4,5}$,
 Le Zhang$^{6}$,
 Emory~F.~Bunn$^{7}$,
 Andrei Korotkov$^{1}$,
 Peter Timbie$^{6}$,
 Gregory~S.~Tucker$^{1}$,
 and Benjamin~D.~Wandelt$^{3,4,2,8}$\\
{~}\\
$^{1}$Department of Physics, Brown University, 182 Hope Street, Providence, RI 02912, USA\\
$^{2}$ Department of Physics, 1110 W Green Street, University of Illinois at Urbana-Champaign, Urbana, IL 61801, USA\\
$^{3}$ UPMC Univ Paris 06, UMR 7095, Institut d'Astrophysique de Paris, 98 bis, boulevard Arago, 75014 Paris, France\\
$^{4}$ CNRS, UMR 7095, Institut d'Astrophysique de Paris, 98 bis, boulevard Arago, 75014 Paris, France\\
$^{5}$ Center for Cosmology and Astro-Particle Physics, Ohio State University, Columbus, OH 43210, USA\\
$^{6}$Department of Physics, University of Wisconsin, Madison, WI 53706, USA\\
$^{7}$ Physics Department, University of Richmond, Richmond, Virginia 23173, USA\\
$^{8}$ Department of Astronomy, University of Illinois at Urbana-Champaign, Urbana, IL 61801, USA
}

\thanks{Email:  ata$\_$karakci@brown.edu}


%
%

\begin{abstract}
Detection of $B$-mode polarization of the cosmic microwave background (CMB) radiation is one of the frontiers of observational cosmology. Because they are an order of magnitude fainter than $E$-modes, it is quite a challenge to detect $B$-modes. Having more manageable systematics, interferometers prove to have a substantial advantage over imagers in detecting such faint signals. Here, we present a method for Bayesian inference of power spectra and signal reconstruction from interferometric data of the CMB polarization signal by using the technique of Gibbs sampling. We demonstrate the validity of the method in the flat-sky approximation for a simulation of an interferometric observation on a finite patch with incomplete $uv$-plane coverage,  a finite beam size and a realistic noise model. With a computational complexity of $O(n^{3/2})$, $n$ being the data size, Gibbs sampling provides an efficient method for analyzing upcoming cosmology observations.
\begin{center}
\emph{Subject headings}: cosmic background radiation - cosmology:observations - instrumentation:interferometers - methods: data analysis - methods: statistical - techniques: polarimetric
\end{center}
\end{abstract}

\section{Introduction}
The cosmic microwave background (CMB) polarization signal can be decomposed into a scalar $E$ component and a pseudo-scalar $B$ component (Zaldarriaga \& Seljak 1997; Kamionkowski et al. 1997). The largest contribution to the CMB polarization comes from the scalar metric perturbations produced by density fluctuations, which produce only $E$-type polarization. At small angular scales ($\ell \sim 1000$) gravitational lensing due to large-scale structure transforms a small portion of the $E$-modes into $B$-modes (Zaldarriaga \& Seljak 1998). The more interesting source of $B$-type polarization is the primordial tensor metric perturbations produced by gravitational waves created during inflation (Zaldarriaga \& Seljak 1997; Kamionkowski et al. 1997). Since tensor modes dominate on large angular scales ($\ell \sim 100$), detection of $B$-modes at these scales offers an excellent probe for the inflationary epoch whose energy scale is proportional to the amplitude of primordial gravitational waves (Hu \& White 1997). 

Because $B$-modes are not produced by scalar perturbations, they are smaller than $E$-modes by more than an order of magnitude. Detection of such weak signals, at a level of tensor-to-scalar ratio of $0.01$, requires excellent control of systematic effects. Since traditional imagers measure $Q$ and $U$ Stokes parameters by differencing two orthogonal polarizations, mismatched beams and pointing errors cause leakage from the much stronger temperature signal into the $Q$ and $U$ signals, significantly contaminating the much weaker $B$-modes (Hu et al. 2003). Interferometers, on the other hand, directly measure the Stokes parameters without subtraction of the signals from different detectors. Thus, mismatch in the beam patterns or differential pointing errors do not cause contamination of the polarization by the temperature signal (Bunn 2007). Morover, for finite sky patches and pixellated maps, $E$ and $B$-modes mix into each other, causing major contamination of $B$-modes by much stronger $E$-modes (Lewis et al. 2002; Bunn 2003). Since interferometric data live in Fourier space, separation of $E$ and $B$-modes can be achieved more cleanly by interferometers than imagers (Park et al. 2003;  Park \& Ng 2004).

Interferometers have already been applied to the detection of the polarized CMB signal. The first detection of polarization anisotropies in the CMB was achieved by DASI (Kovac et al. 2002). CBI (Pearson et al. 2003) and VSA (Dickinson et al. 2004; Grainge et al. 2003) obtained detailed observations enabling them to extract the $E$-mode polarization angular power spectrum up to $\ell \sim 600$. Advancing techniques, such as bolometric interferometry employed by the QUBIC experiment (Battistelli et al. 2010), provide promising developments in detecting the long-sought $B$-mode polarization anisotropies yet to be observed.

In comparison to alternative methods of extracting power spectra such as maximum likelihood and pseudo-$C_l$ estimators, which often scale as $O(n^3)$ and $O(n^{3/2})$ respectively, the method of Gibbs sampling (Jewell et al. 2004; Wandelt et al. 2004) has an advantage in dealing with the demands of future cosmology observations because it provides simultaneous estimation of power spectrum and signal by sampling them from the joint posterior probability density through a Markov chain Monte Carlo process with $O(n^{3/2})$ computational complexity. Gibbs sampling has already been used to analyze the WMAP temperature (O'Dwyer et al. 2004; Dickinson et al. 2009; Larson et al. 2011) and polarization (Larson et al. 2007; Eriksen et al. 2007; Komatsu et al. 2011) data. Sutter et al. (2012) examined the application of Gibbs sampling to interferometric observations of the CMB temperature signal.

Here, we will investigate the application of Gibbs sampling to the polarized CMB signal observed by an interferometer. Our analysis is an extension of Gibbs sampling as applied to interferometric data (Sutter et al. 2011) to polarized signals along the lines of Larson et al. (2007). In Section 2 we investigate the method of Gibbs sampling as applied to interferometric polarimetry. In Section 3. we discuss the simulation of interferometric observation of the polarized CMB signal on a finite patch in the flat-sky approximation. In Section 4 we present our results of polarized power spectra and signal reconstructions. Finally, in Section 5 we make some comments and concluding remarks. 

\section{Method of Gibbs Sampling}

In the flat-sky approximation, we describe the CMB signal as a $3 n_p$ dimensional vector, $\vec s$, of the Fourier transform of the discretized sky maps of $n_p$ pixels; $\vec{s} = (\ldots, \tilde T_i, \tilde E_i, \tilde B_i, \ldots)$;  $i = 0, \ldots, n_p - 1$, where $\tilde f$ denotes the Fourier transform of $f$. The covariance matrix $\mathbf S = \left< \vec s ~ \vec s^{~\dagger}  \right>$ of the CMB signal is a block diagonal matrix with a $3 \times 3$ submatrix $\mathbf C_i$ at each pixel $i$:

\begin{equation} \label{eq:cl} 
\mathbf C_i = \left( \begin{array}{ccc}
C_{\ell_i}^{TT} & C_{\ell_i}^{TE} & C_{\ell_i}^{TB} \\
C_{\ell_i}^{TE} & C_{\ell_i}^{EE} & C_{\ell_i}^{EB} \\
C_{\ell_i}^{TB} & C_{\ell_i}^{EB} & C_{\ell_i}^{BB} \end{array} \right)
\end{equation}
where $\ell_i = 2\pi |\vec u_i|$ and $\vec u_i$ is the position vector of the $i^{th}$ pixel in the Fourier plane. Larson et al. (2007) have applied Gibbs sampling to the full-sky WMAP polarization data. Their analysis can be extended to interferometric observations by describing the visibility data, $\vec d$, from a polarimetric observation as pixelated maps of the Stokes parameters $T, Q,$ and $U$:

\begin{equation} \label{eq:dat}
\vec{d} = \mathbf H ~ \mathbf R ~ \vec{s} + \vec{n}
\end{equation}
where  $\vec n$ is a Gaussian realization of the noise,  $\mathbf H$ is a linear operator that includes convolution with an instrument beam $A$ and $\mathbf R$, a block diagonal matrix with a $3 \times 3$ submatrix $\mathbf R_i$ at each pixel, is the transformation of the $\tilde T$, $\tilde E$, and $\tilde B$ components of the signal $\vec s$ into the Fourier transform of the Stokes parameters. For an interferometric data set, in the flat-sky approximation, $\mathbf H$  and $\mathbf R_i$ can be written as

\begin{equation} \label{eq:eyc}
\mathbf H = I F A F^{-1},
\end{equation}

\begin{equation} \label{eq:ri}
\mathbf R_i = \left( \begin{array}{ccc}
1 & 0 & 0 \\
0 & cos(2\phi_i) & -sin(2\phi_i) \\
0 & sin(2\phi_i) & cos(2\phi_i) \end{array} \right)
\end{equation}
where $F$ is a Fourier transform operator, $I$ is an interferometer patten in the $uv$-plane, and $\phi_i$ is the angular position of the $i^{th}$ pixel in the $uv$-plane.

Larson et al. (2007) investigated Gibbs sampling as applied to polarized signals. The principle approach is to sample $\vec s$ and $\mathbf S$ from the joint density $P(\mathbf S, ~\vec s, ~\vec d~)$, which can be obtained by a Markov chain Monte Carlo process by successively sampling from the conditional distributions $P(\vec s ~ | ~ \mathbf S, ~\vec d~)$ and $P(\mathbf S ~ | ~ \vec s, ~\vec d~) \propto P(\mathbf S ~ | ~ \vec s~)$. Starting from an initial guess $\mathbf S^0$, sampling is done in an iterative fashion by (Larson et al. 2007)

\begin{equation} \label{eq:sis}
\vec s^{~a+1} \gets P(\vec s ~ | ~ \mathbf S^a, ~\vec d~),
\end{equation}

\begin{equation} \label{eq:cis}
\mathbf S^{a+1} \gets P(\mathbf S ~ | ~ \vec s^{~a+1} ).
\end{equation}
After some ``burn-in" steps the stationary distribution of the Markov chain is reached and the samples approximate to being samples from the sought-after joint distribution. Sampling from the joint distribution by this technique is called the method of \emph{ Gibbs sampling}.

Given the current covariance matrix $\mathbf S^a$, the sky signal, $\mathbf R ~ \vec s^{~a+1} = \vec x + \vec y$, separated into the mean field, $\vec x$, and fluctuation, $\vec y$, parts, is sampled by solving the following equations (Larson et al. 2007):

\begin{equation} \label{eq:mean}
(\mathbf R ~ \mathbf S^{-1} \mathbf R^T + \mathbf H^T \mathbf N^{-1} \mathbf H) \vec x = \mathbf H^T \mathbf N^{-1} \vec d ,
\end{equation}

\begin{equation} \label{eq:fluc}
(\mathbf R ~ \mathbf S^{-1} \mathbf R^T + \mathbf H^T \mathbf N^{-1} \mathbf H) \vec y = \mathbf R ~ \mathbf S^{-1/2} \vec \xi + \mathbf H^T \mathbf N^{-1/2} \vec \chi
\end{equation}
where $\mathbf N = \left<\vec n ~ \vec n^{~\dagger} \right>$ is the noise covariance matrix, which is diagonal (White et al. 1999) and has entries equal to $N_{ij} = \nu^2_i \delta_{ij}$ where $\nu_i$ is the noise variance for the $i^{th}$ pixel in the $uv$-plane, and $\vec \xi$ and $\vec \chi$ are Gaussian random maps having values of zero mean and unit variance in each pixel for each of the $T, Q$, and $U$ components.

We obtain numerical solutions for the Eq. \ref{eq:mean} and Eq. \ref{eq:fluc} by the preconditioned conjugate-gradient method (Press et al. 1986). A good choice for the preconditioner is the inverse of the diagonal part of the operator:  

\begin{equation} \label{eq:preq}
\mathbf {P}^{-1} =  \mathbf{P}_{N}^{-1} + \emph{diag}\{\mathbf R \mathbf S^{-1} \mathbf R^T \}.
\end{equation}
The noise portion of the preconditioner, $\mathbf{P}_{N}$, can be written as (Sutter et al. 2012)

\begin{equation} \label{eq:preno}
\mathbf{P}_{N}^{-1} = F(F^{-1} I (I \mathbf N I)^{-1}I)(F^{-1} \tilde{A}^2)
\end{equation}
where $\tilde{A}$ is the Fourier transform of the beam pattern.

The signal polarization map, $\vec s^{~a}$, sampled from $P(~\vec s ~|~ \mathbf S^{a-1}, ~\vec d~)$, is used to sample the signal covariance matrix from $P(~\mathbf S ~|~ \vec s^{~a} ~)$ by computing the unnormalized variance $\sigma _\ell$ in an annulus of radius $\ell / 2 \pi$. We can define uniform bins  $b = [\ell_{min}, ~\ell_{max}]$ in which $C_\ell \ell(\ell+1)$ is roughly constant. Then $\sigma_\ell$ is defined for bin $b$ as (Larson et al. 2007)

\begin{equation} \label{eq:var}
\sigma_b = \sum_{\ell_i \in b} \ell_i(\ell_i +1) \vec s_i ~  \vec s_i ^{~\dagger}
\end{equation}
where $\vec s_i = (\tilde T_i, \tilde E_i, \tilde B_i)$ is a three-vector at the $i^{th}$ pixel. Sampling from the probability density $P~(~\mathbf C_b ~|~ \vec s^{~a} ~)$, which is an inverse Wishart distribution with $m_b$ degrees of freedom, can be done by drawing $m_b = p_b - 2$ (assuming a Jeffreys' ignorance prior) vectors from a Gaussian distribution with covariance matrix $\sigma^{-1}_b$, where $p_b$ is the number of pixels in the bin $b$. The required sample $\mathbf C_b$ is, then, the inverse of the sum of outer products of these independently sampled vectors (Larson et al. 2007). The actual power spectrum coefficients are given by $\mathbf C_\ell = \mathbf C_b / \ell (\ell + 1)$.

Following Sutter et al. (2012), the Gelman-Rubin (G-R) statistic is employed to determine that the stationary distribution of the Markov chain has been reached. Given multiple instances of chains,  convergence is reached when the \emph{potential scale reduction factor} of the G-R statistic, determined by the ratio of the variance within each chain to the variance among chains, assumes a value less than a given tolerance for each bin (Gelman \& Rubin 1992).

\section{Simulations}

We construct the input $Q$ and $U$ maps by transforming realizations of $E$ and $B$ signals over 10-degree square patches with $128$ pixels per side. The realizations are created as maps of Gaussian fluctuations with a covariance $\mathbf C_i $ (Eq. 1)  at each pixel whose components are produced by CAMB (Lewis et al. 2000). The cosmological parameters used for CAMB are consistent with the 7-year results of WMAP (Larson et al. 2011; Komatsu et al. 2011); $\Omega_M = 0.27$, $\Omega_\Lambda = 0.73$, $\Omega_b = 0.045$, and $H_0 = 70km s^{-1} Mpc^{-1}$. The tensor-to-scalar ratio is taken to be $T/S = 0.01$.  The angular resolution of the signal maps is 4.7 arcminutes corresponding to a maximum available multipole of $\ell_{max} = 2\,300$. The spatial resolution in the $uv$-plane is 5.73 $\lambda$. Although the patch size is too large to employ the flat-sky approximation, it is still useful in exploring the validity of our technique.

The primary beam pattern $A$ is modeled as a Gaussian with peak value of unity and standard deviation of 2.5 degrees allowing us to include all Fourier modes up to the Nyquist frequency in the analysis. Although a smaller beam size would further reduce the edge-effects caused by the periodic boundary conditions of the fast Fourier transformations, it would also require a longer computation time.

The interferometer is constructed by randomly placing 16 antennas with diameters of 10 cm in the $uv$-plane and uniformly rotating the baselines over a period of 12 hours while observing the same sky patch. The observation frequency is 30 GHz with a 10-GHz bandwidth. With this frequency and antenna radius the minimum available multipole is $\ell_{min} = 28$. The interferometer pattern $I$ is constructed by placing a value of one at each pixel that coincides with a baseline length during the observation period and zeros everywhere else. With the given number of antennas and pixel resolution, the resulting interferometer pattern provides us with a fairly realistic case of incomplete $uv$-plane coverage as shown in Figure 1. With this configuration, the $uv$-plane coverage is roughly 70\%, which varies for each $\ell$-bin, as shown in Figure 2.

\begin{figure} \label{fig:inter}
\begin{center}
    \leavevmode
    \includegraphics[trim = 2.2cm 0.5cm 2.5cm 1.4cm, clip=true, width=8cm]{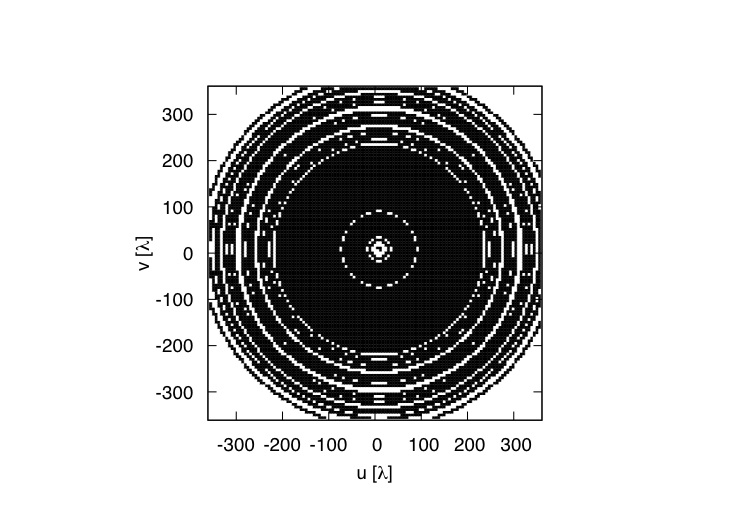}
    \caption{Interferometer pattern, $I$, created over an observation period of 12 hours by 16 antennas of radius $5 \lambda$ randomly placed in the $uv$-plane.}
\end{center}
\end{figure}

The noise at each pixel for the temperature data is obtained from the total observation time that all baselines spend in the pixel. The noise variance is given as $\nu_i \propto 1 /  \sqrt{t^{obs}_i}$. The overall temperature signal-to-noise ratio is set to 50 by scaling the noise variance at each pixel by the constant $| \mathbf H ~ \mathbf R ~ \vec{s} | / 50| \vec{n} |$. Gaussian realizations of this noise are used to create the $T, Q$, and $U$ data. The corresponding signal-to-noise ratios for $Q$ and $U$ signals become 2.2 and 2.3 respectively.

\begin{figure} \label{fig:cover}
\begin{center}
    \leavevmode
    \includegraphics[width=8.7cm]{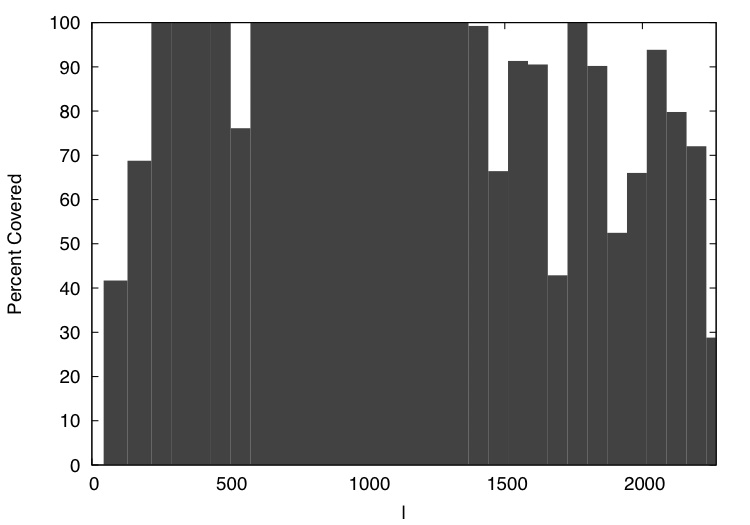}
    \caption{Coverage of $uv$-plane for each bin $\Delta_{b} = [\ell_{b}^{min}, \ell_{b}^{max}]$. Shown are the percentages of pixels intersected by baseline vectors during the 12-hour observation period in each bin. A pixel in Fourier space is said to be in bin $b$ if its position vector $\vec u_{pix}$ satisfies $2\pi |\vec u_{pix}| \in \Delta_{b}$.}
\end{center}
\end{figure}

\begin{figure} \label{fig:cong}
\begin{center}
    \leavevmode
    \includegraphics[width=8.7cm]{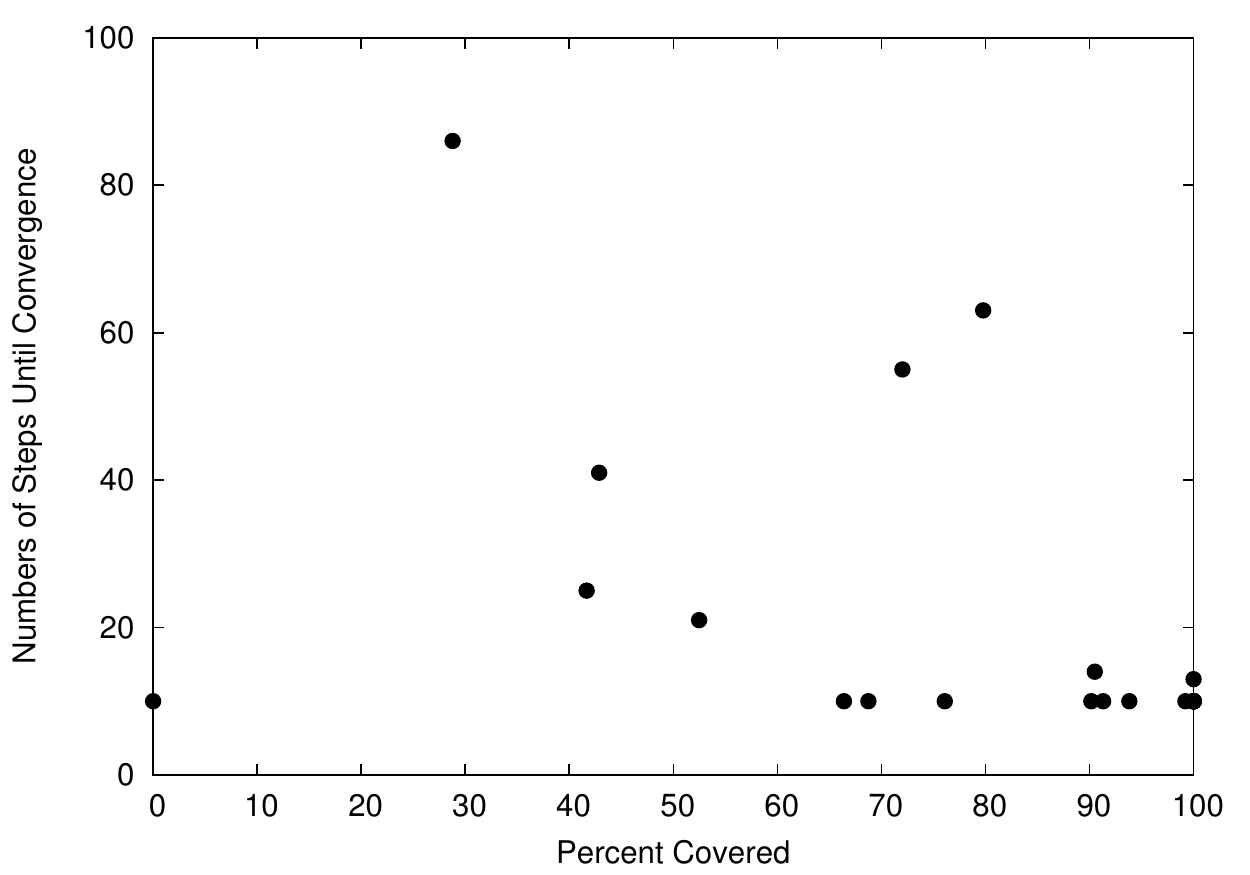}
    \caption{The percentage of $uv$-plane coverage versus the number of steps, after the burn-in phase, required to reach convergence at each $\ell$-bin.}
\end{center}
\end{figure}

Our analysis had four independent chains, each chain having $2\,200$ iterations of which the first $200$ are discarded for a ``burn-in" phase. The G-R statistic reached less than 1.2 \footnote[1]{Gelman et al. (2004) suggests that values below 1.2 are ``acceptable."} for each bin in about 83 hours. The computation spent 150MB of memory on 4 cores of an Intel dual six-core X5650 Westmere 2.66GHz machine.

Figure 3 shows the $uv$-plane coverage versus the number of steps, after burn-in, required to reach convergence for each bin. We see that convergence time and $uv$-plane coverage are weakly correlated. Incomplete coverage leads to a larger correlation length for small power leading to a longer convergence time. Therefore, low-coverage bins have larger effect on overall performance, as expected.

\section{Results}

\begin{figure*} \label{fig:spect}
  \begin{center}$
    \leavevmode
    \begin{array}{c@{\hspace{1.5cm}}c}
     \includegraphics[trim = 1mm 1mm 1mm 1mm, width=7.5cm]{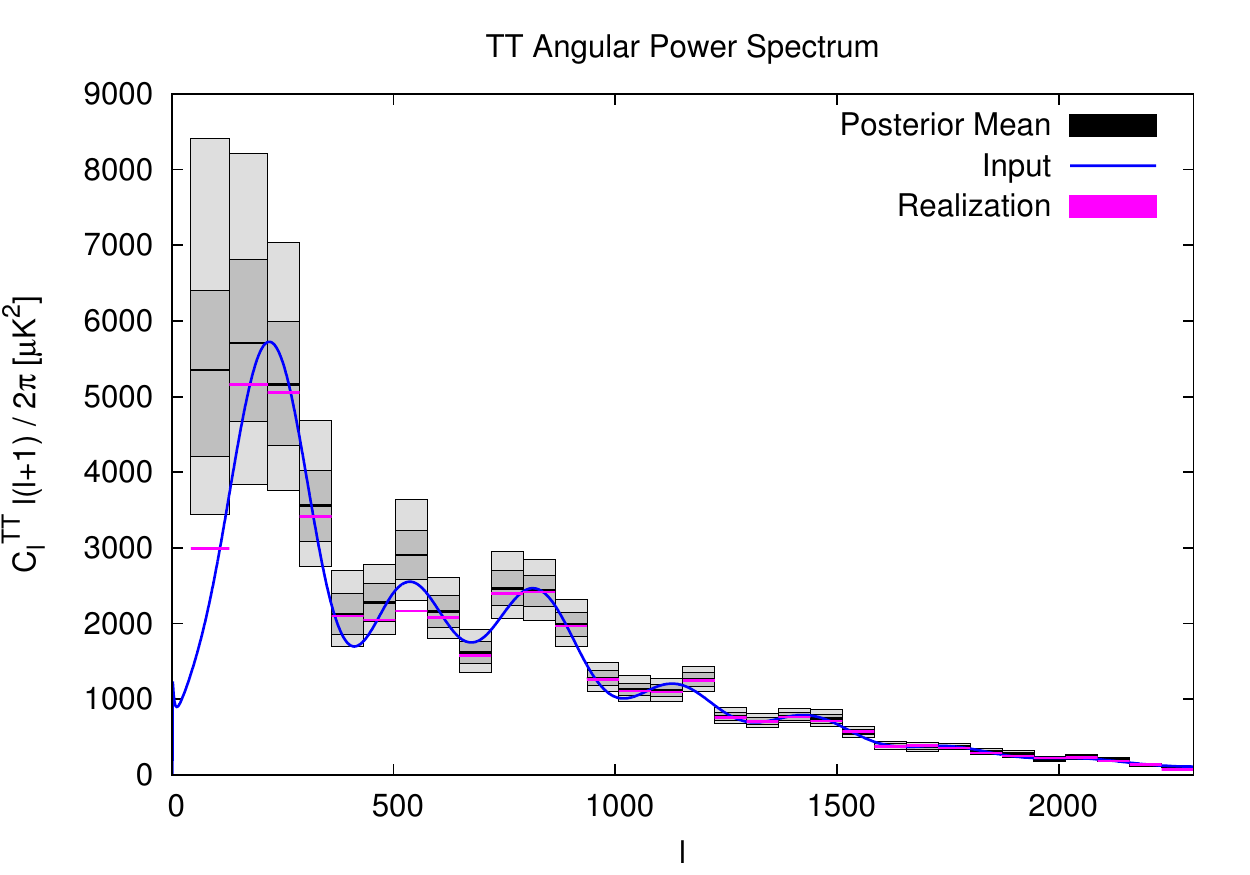} &
     \includegraphics[trim = 1mm 1mm 1mm 1mm, width=7.5cm]{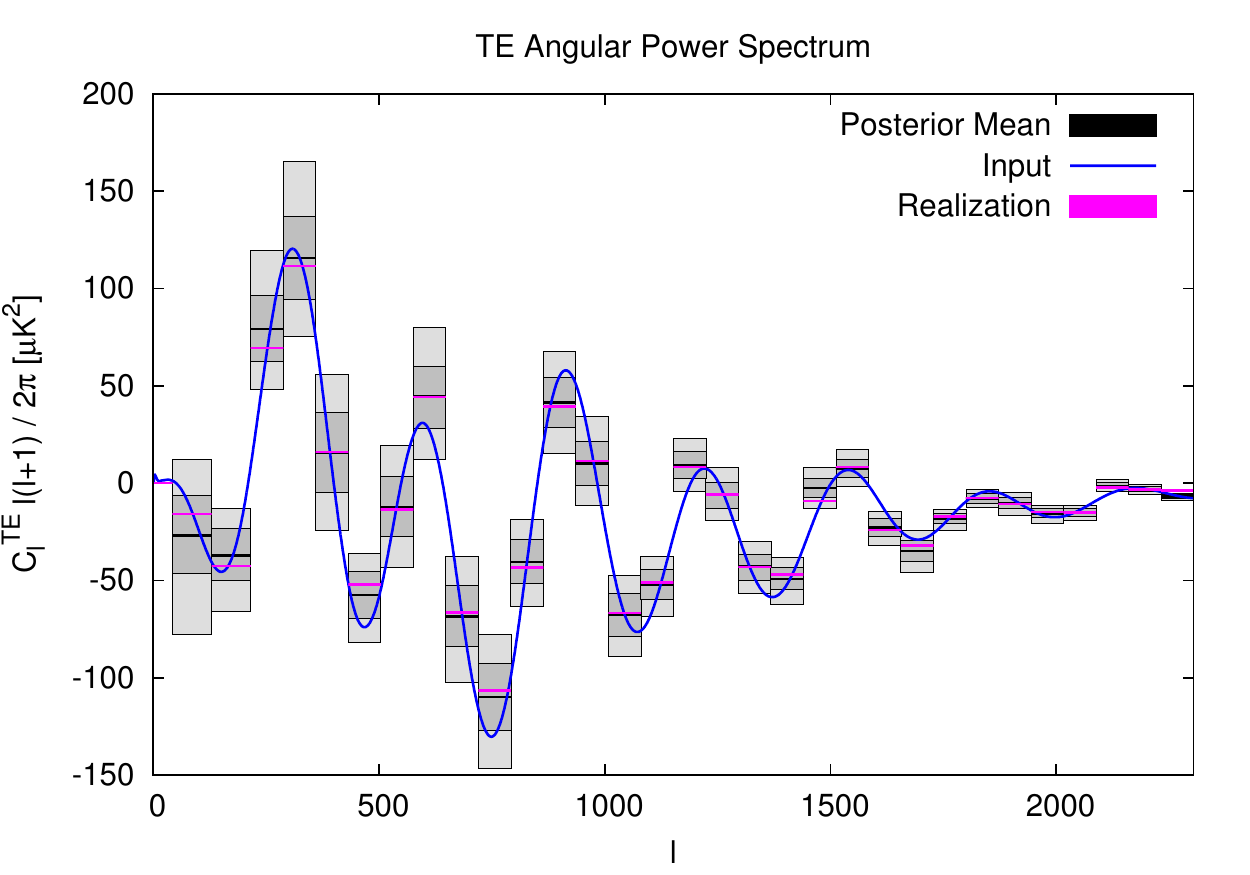} \\
     \includegraphics[trim = 1mm 1mm 1mm 1mm, width=7.5cm]{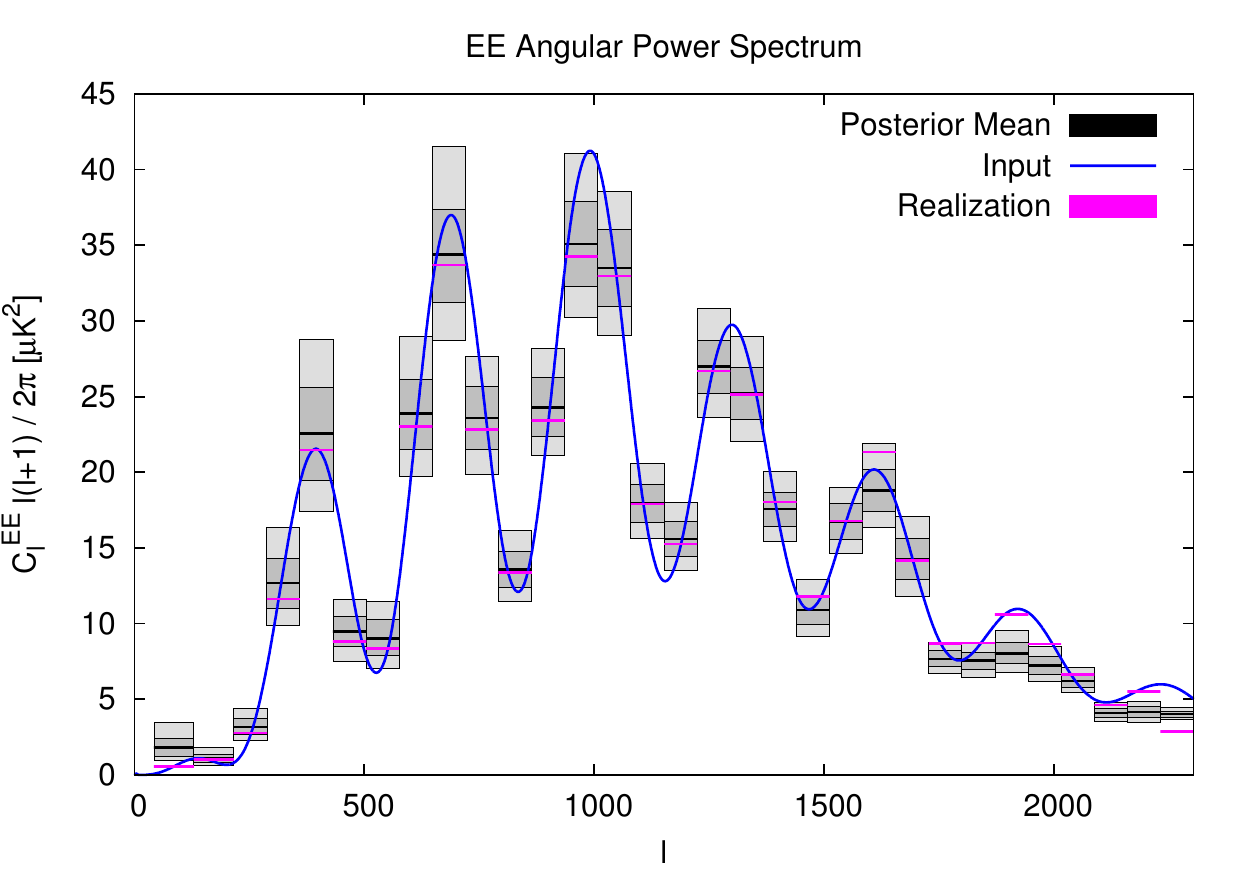} &
     \includegraphics[trim = 1mm 1mm 1mm 1mm, width=7.5cm]{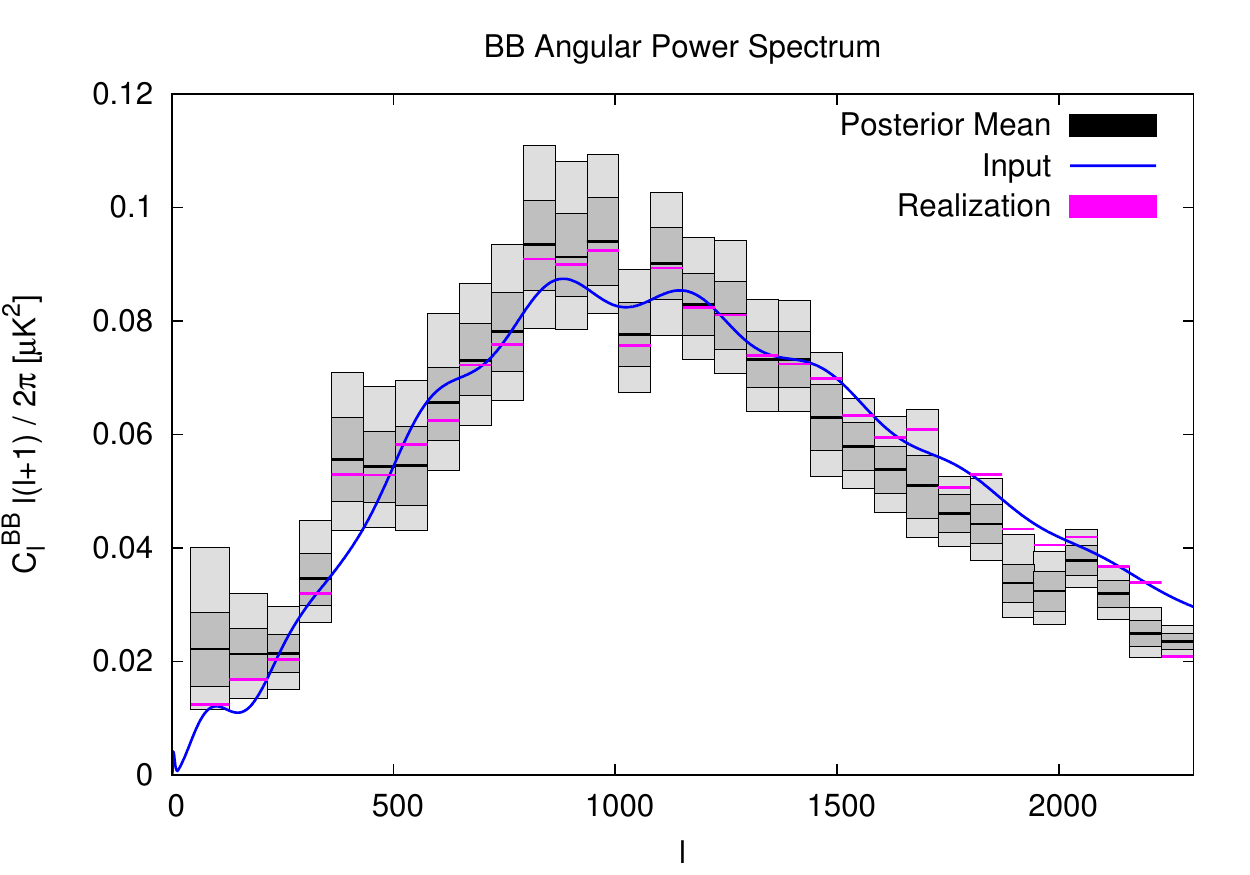} \\
     \includegraphics[trim = 1mm 1mm 1mm 1mm, width=7.5cm]{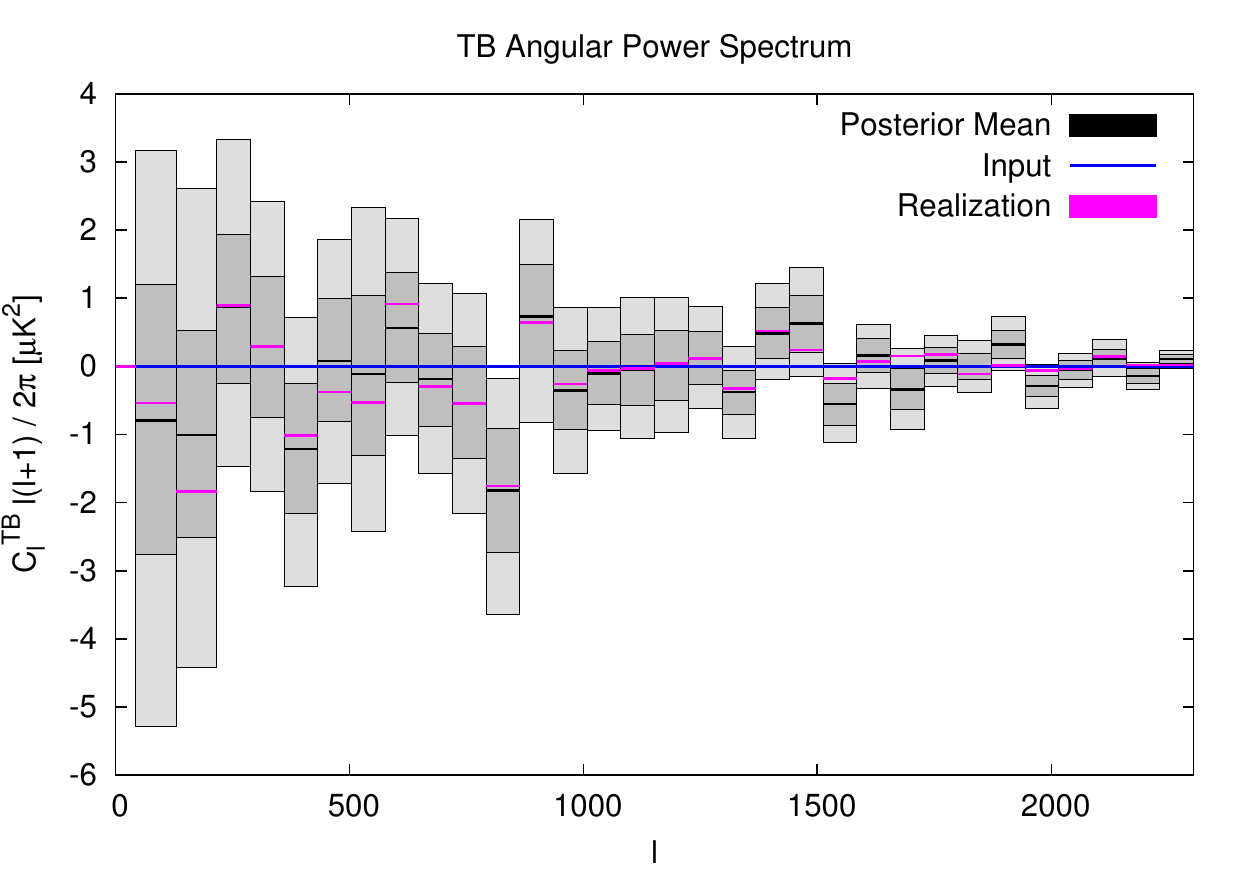} &
     \includegraphics[trim = 1mm 1mm 1mm 1mm, width=7.5cm]{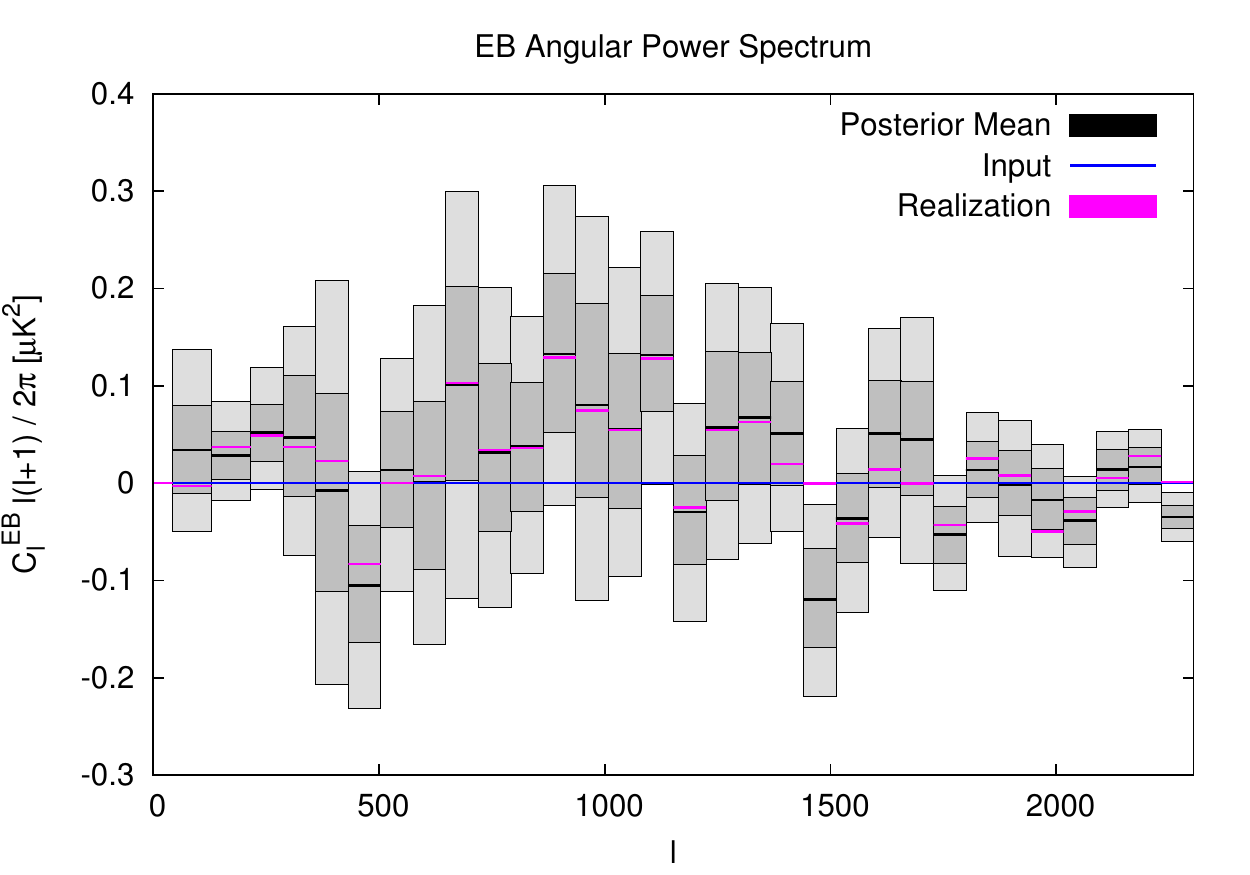} \\
     \end{array}$
       \caption{Mean posterior power spectra for each $\ell$-bin are shown in black. Dark and light grey indicate $1\sigma$ and $2\sigma$ uncertainties, respectively. The binned power spectra of the signal realization are shown in pink. Blue lines are the input CMB power spectra obtained by CAMB for a tensor-to-scalar ratio of $T/S = 0.01$.}
  \end{center}
\end{figure*}

\subsection{Power Spectra}

The mean posterior power spectra of the four independent chains, together with the associated uncertainties at each $\ell$-bin obtained after convergence is reached, is shown in Figure 4. The input power spectra, which are used to construct signal realization, and the spectra of the signal realization are also shown in Figure 4. Nearly all of our estimates fall within $1\sigma$ of the expected value, and none of them is outside of the $2\sigma$ width.

Although the effect of the incomplete $uv$-plane coverage is not evident in Figure 4, if we consider  the sizes of the uncertainties, relative to the mean posterior, we see that at the bins with weak $uv$-plane coverage the relative sizes of the uncertainties are larger whereas at the bins with complete coverage the relative sizes are smaller as expected. The effect of sample variance dominates at low $\ell$ values where the sizes of the uncertainties, relative to the mean posterior, are larger due to the finite size of the sky patch.

Gibbs sampling also provides higher-order statistical information such as the two-point correlations between $\ell$-bins. Off-diagonal components of the correlation matrices for $TE$, $EE$ and $BB$ power spectra are shown in Figure 5. There is a slight correlation between adjacent bins, which is the result of reduced Fourier space resolution caused by finite beam width. The correlation is more pronounced at high $\ell$ and low signal-to-noise ratio, as seen in $BB$ correlation matrix of Figure 5. Since the correlation matrices carry information about data regions larger than bin sizes, the power spectra are oversampled causing anti-correlation between nearby bins. Since the power in a region is constraint by the data, whenever a large value is sampled at a certain bin values of samples from the other bins in the same data region are reduced (Elsner \& Wandelt 2012).

\begin{figure*} \label{fig:cormat}
  \begin{center}$
    \leavevmode
    \begin{array}{ccc}
     \includegraphics[trim = 2cm 1cm 2cm 1cm, clip=true, width=5.5cm]{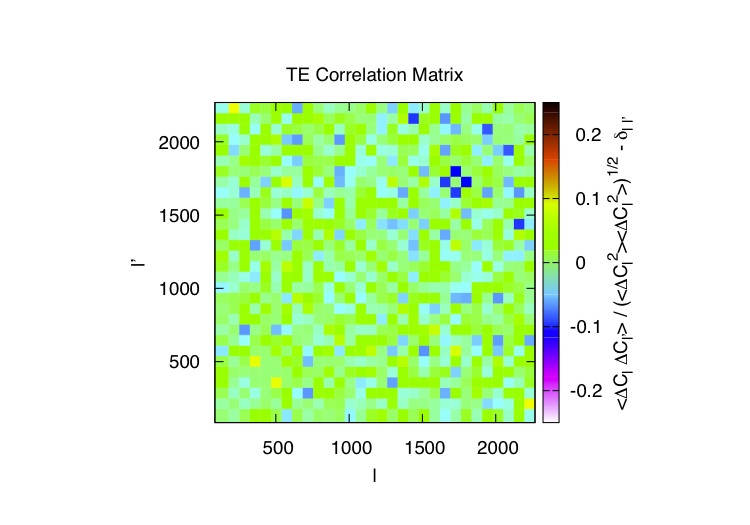} &
     \includegraphics[trim = 2cm 1cm 2cm 1cm, clip=true, width=5.5cm]{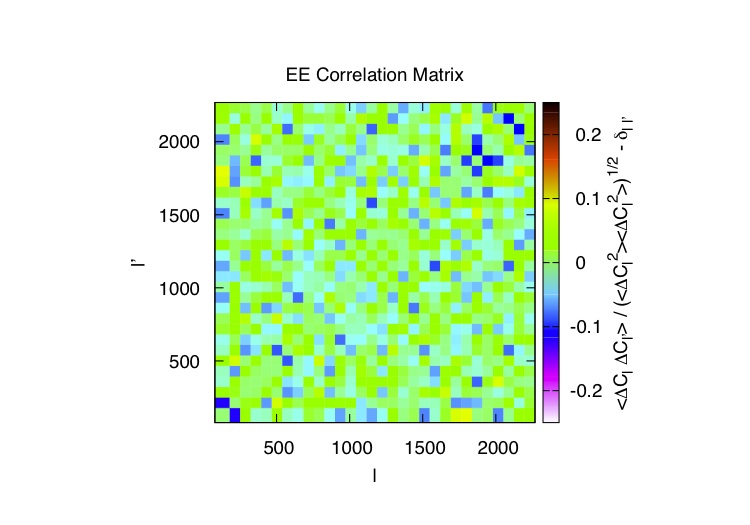} &
     \includegraphics[trim = 2cm 1cm 2cm 1cm, clip=true, width=5.5cm]{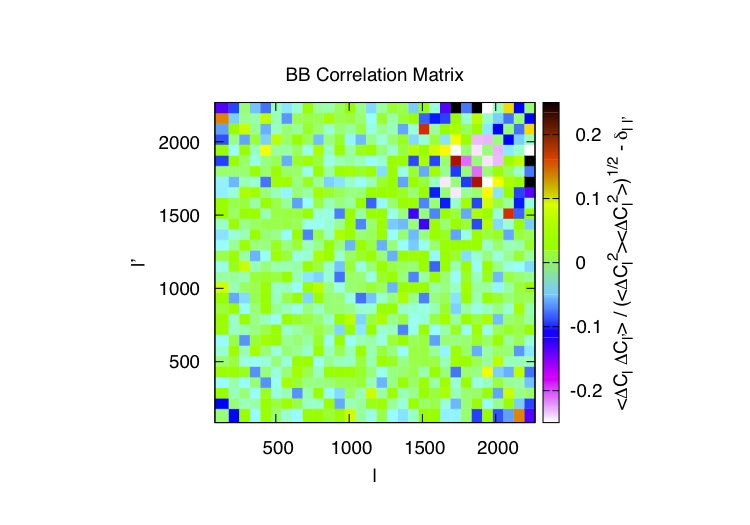} \\
     \end{array}$
       \caption{Correlation matrices of $TE$, $EE$ and $BB$ power spectra $-$ only the off-diagonal elements are shown. Correlations and anti-correlations between nearby power spectrum bins are the results of having a finite beam width and a finite bin size, respectively. Correlations are stronger towards lower signal-to-noise values (from $TE$ to $BB$) and towards higher $\ell$-values.}
  \end{center}
\end{figure*}

\begin{figure*} \label{fig:hitop}
\begin{center}$
\leavevmode
\begin{array}{ccccc}
\includegraphics[trim = 1mm 1mm 7.1mm 1mm, clip=true, width=6.2cm]{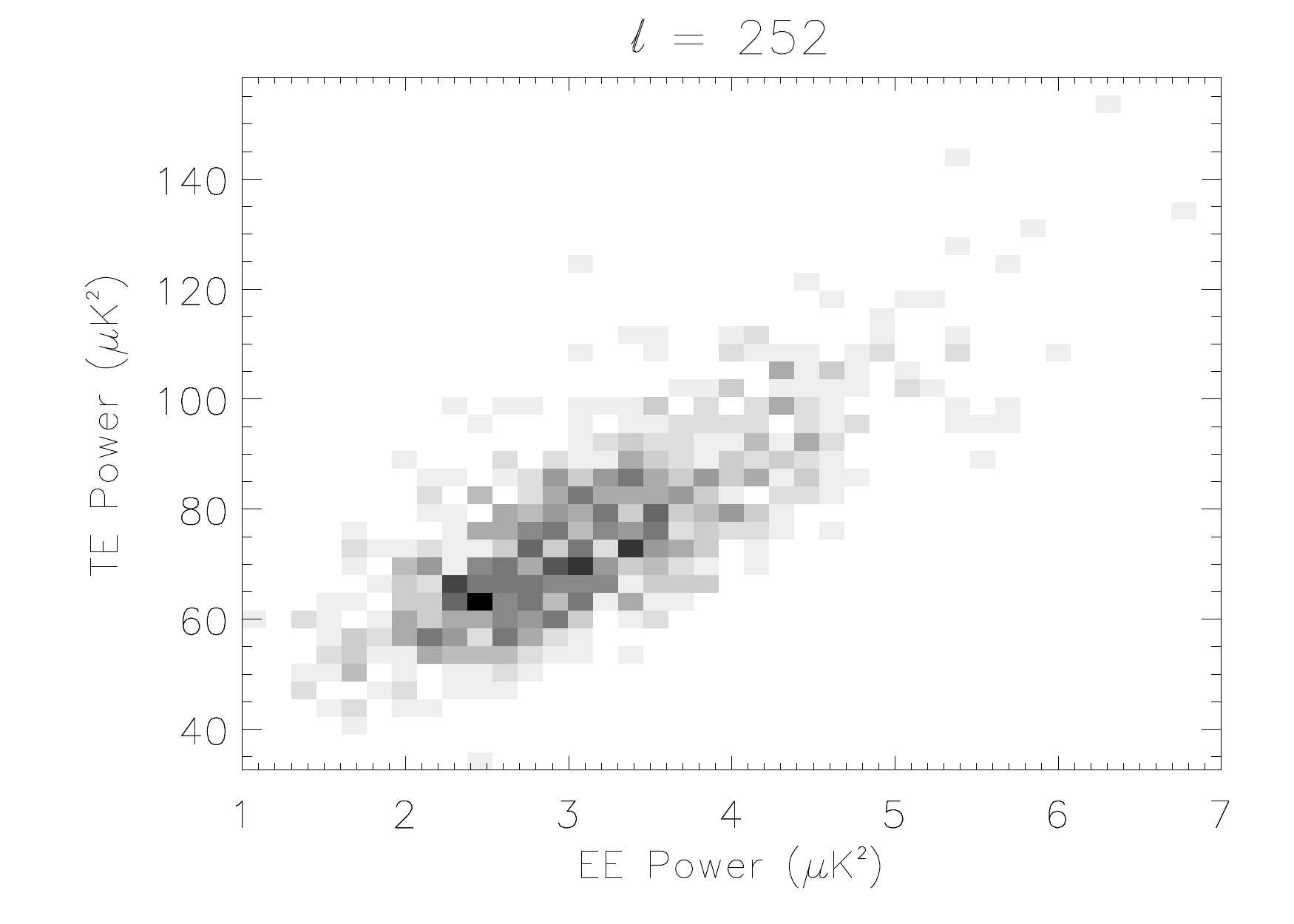} &
\raisebox{1.3cm}{\includegraphics[trim = 1.5mm 1mm 1mm 1mm, clip=true, width=7.5mm]{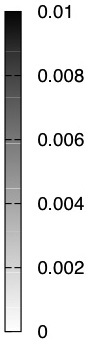}} & \hspace{1cm} &
\includegraphics[trim = 1mm 1mm 7.1mm 1mm, clip=true, width=6.2cm]{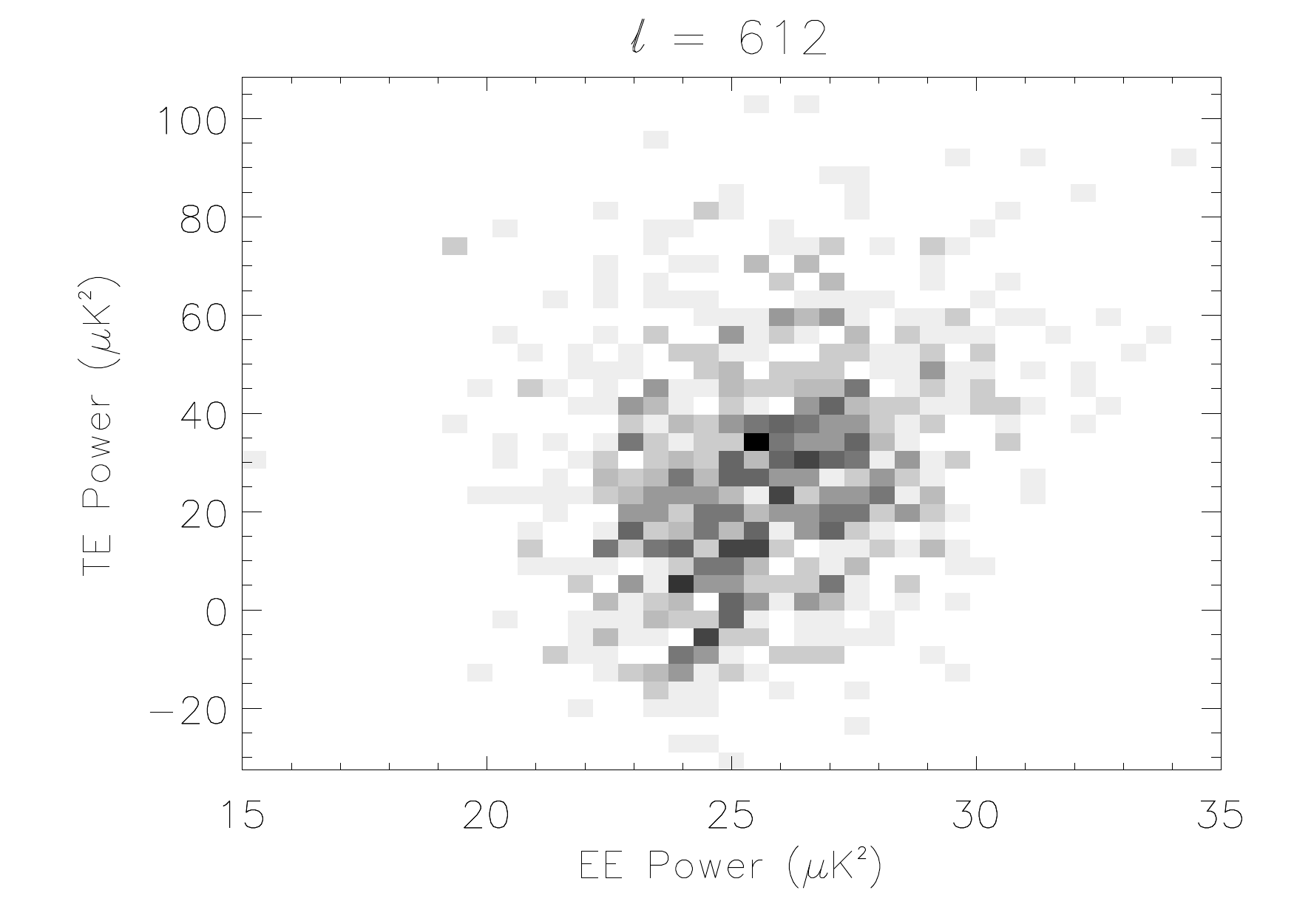} &
\raisebox{1.3cm}{\includegraphics[trim = 1.5mm 1mm 1mm 1mm, clip=true, width=7.5mm]{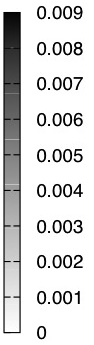}} \\
\includegraphics[trim = 1mm 1mm 7.1mm 1mm, clip=true, width=6.2cm]{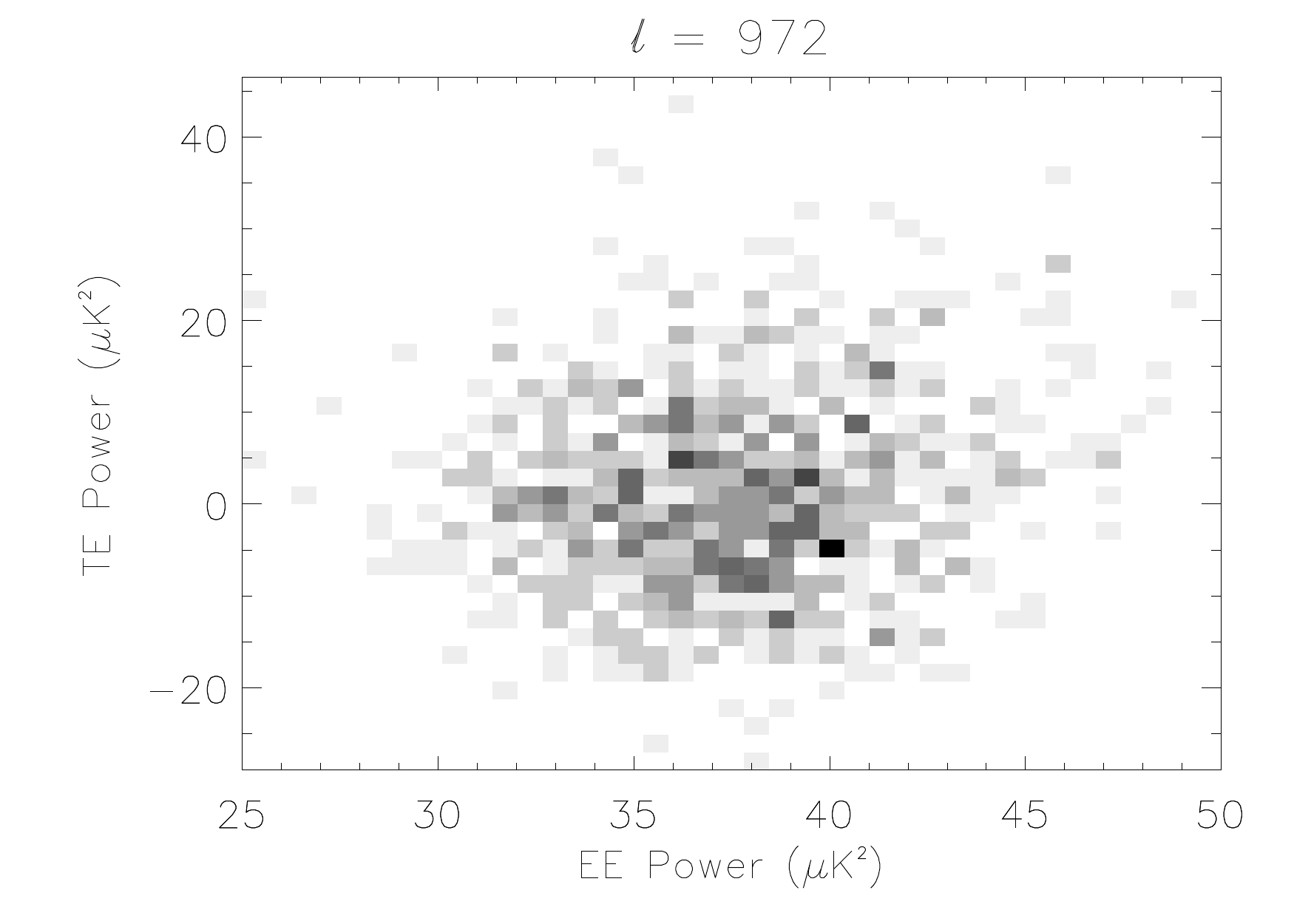} &
\raisebox{1.3cm}{\includegraphics[trim = 1.5mm 1mm 1mm 1mm, clip=true, width=7.5mm]{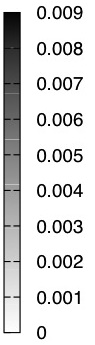}} & \hspace{1cm} &
\includegraphics[trim = 1mm 1mm 7.1mm 1mm, clip=true, width=6.2cm]{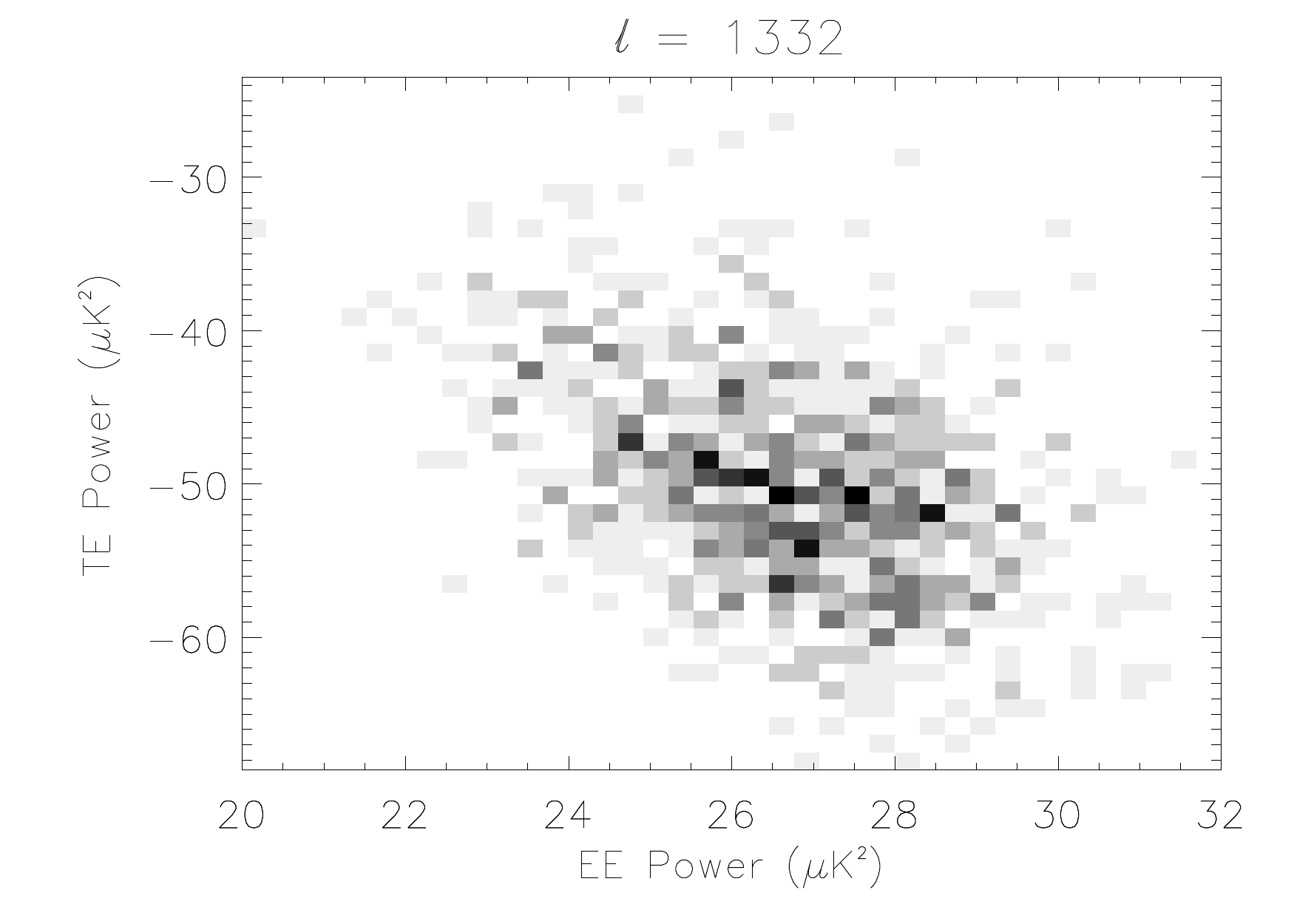} &
\raisebox{1.3cm}{\includegraphics[trim = 1.3mm 1mm 1mm 1mm, clip=true, width=7.5mm]{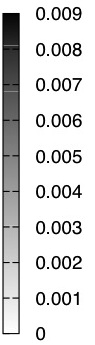}} \\
\includegraphics[trim = 1mm 1mm 7.1mm 1mm, clip=true, width=6.2cm]{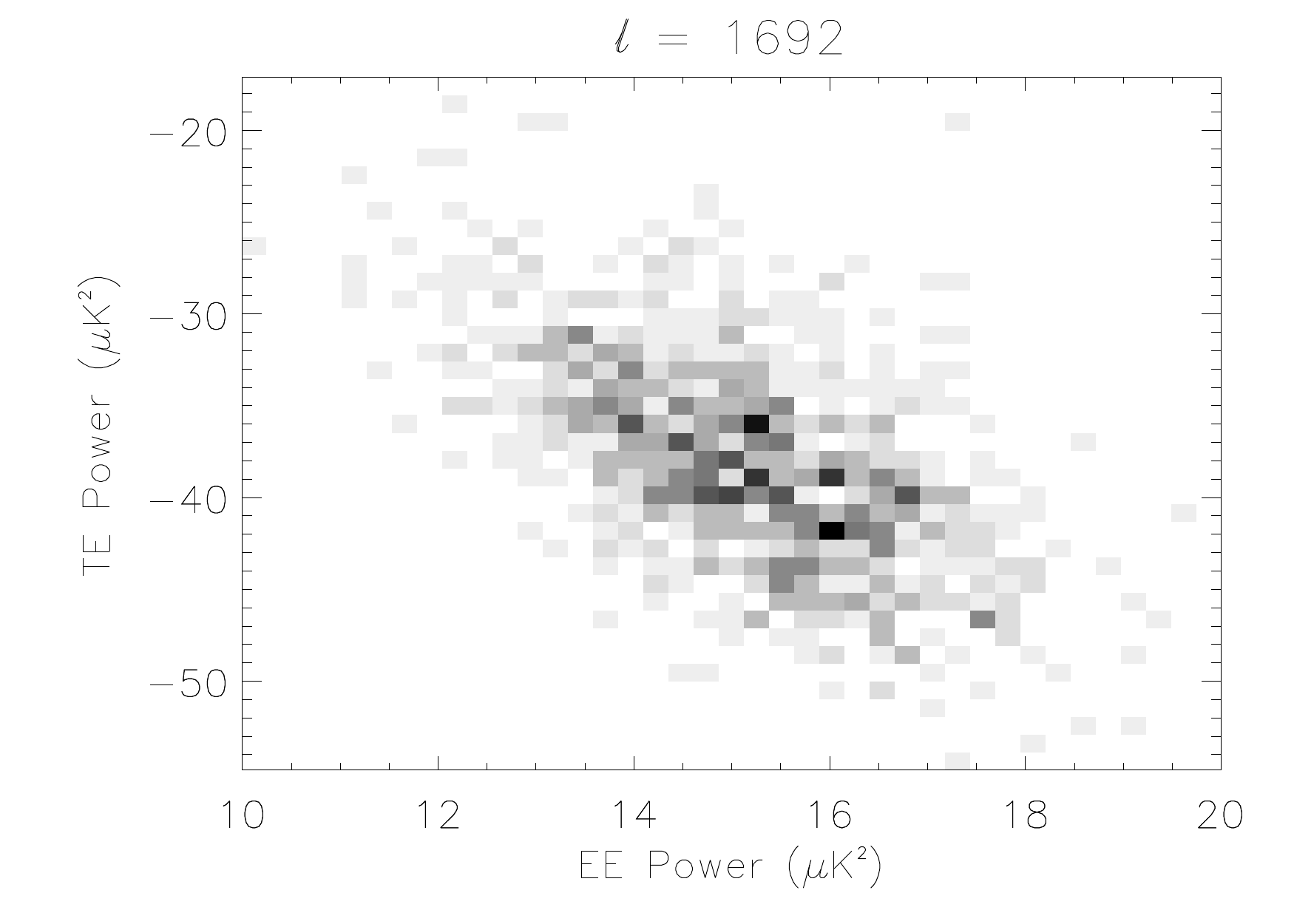} &
\raisebox{1.3cm}{\includegraphics[trim = 1.5mm 1mm 1mm 1mm, clip=true, width=7.5mm]{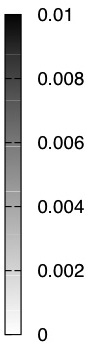}} & \hspace{1cm} &
\includegraphics[trim = 1mm 1mm 7.1mm 1mm, clip=true, width=6.2cm]{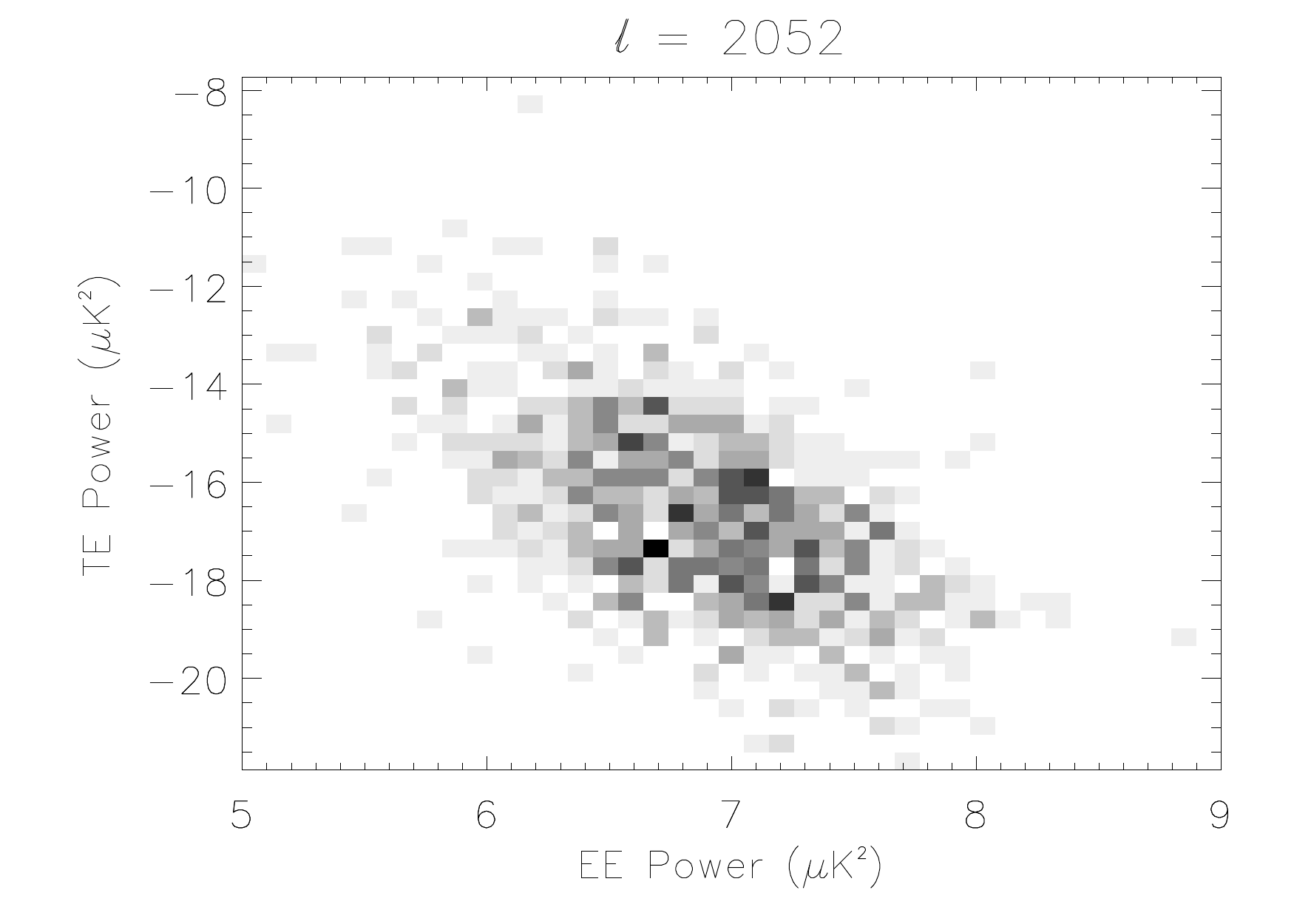} &
\raisebox{1.3cm}{\includegraphics[trim = 1.5mm 1mm 1mm 1mm, clip=true, width=7.5mm]{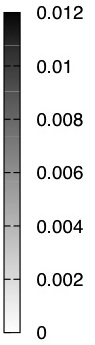}} \\
\end{array}$
\caption{Marginalized posterior joint distributions of $EE$ and $TE$ power spectra for different $\ell$-bins. Samples of power spectra produced by Gibbs sampling have non-Gaussian distributions.}
\end{center}
\end{figure*}

Samples of power spectra produced by Gibbs sampling have highly non-Gaussian probability densities. As an example, we show marginalized posterior joint  distributions of $EE$ and $TE$ power spectra for various $\ell$-bins in Figure 6. Although combining many modes into each bin has an overall Gaussianizing effect, non-Gaussianity of the distributions is still clearly visible. 

\subsection{Signal Reconstruction}

Signal samples are constructed as constrained realizations by adding a fluctuation map to the mean field (Wiener-filtered) map (Hoffman \& Ribak 1991, Bunn et al. 1994, and Elsner \& Wandelt 2012b). We compute Wiener-filtered maps by solving Eq. 7 for $\vec {x}^{a}$. The mean value of these maps $\left<F^{-1}\mathbf{R} \vec{x}\right>$, transformed into Stokes variables $T$, $Q$ and $U$ and averaged over all iterations, are shown in Figure 7. The Wiener filter provides the information content of the data by filtering out the imperfections caused by finite beam, partial $uv$-plane coverage and noise. To obtain a Gaussian random variate for the signal sampling we need to add a fluctuation term with zero mean and the covariance of the conditional posterior, $(\mathbf{S}^{-1} + \mathbf{N}^{-1})^{-1}$, to the mean field map. The fluctuations obtained by solving Eq. 8 provide a random complement to the Wiener-filtered map such that their sum is an unbiased signal sample consistent with the data and the current power spectrum. These artificially created fluctuations average out after sufficient iterations leaving us with a reconstruction of the input signal within the area of the primary beam, which we show in Figure 8 as the ``Final Mean Reconstructed Signal". For comparison, the ``Dirty Map", which is $F^{-1} (\mathbf{H~R} ~\vec s + \vec n)$, is also shown in Figure 8 along with the ``Input Signal" which is constructed from the input power spectra shown in blue in Figure 4.

\begin{figure*} \label{fig:wiener}
  \begin{center}$
    \leavevmode
    \begin{array}{ccc}
     \includegraphics[trim = 2cm 1cm 2cm 1cm, clip=true, width=5.5cm]{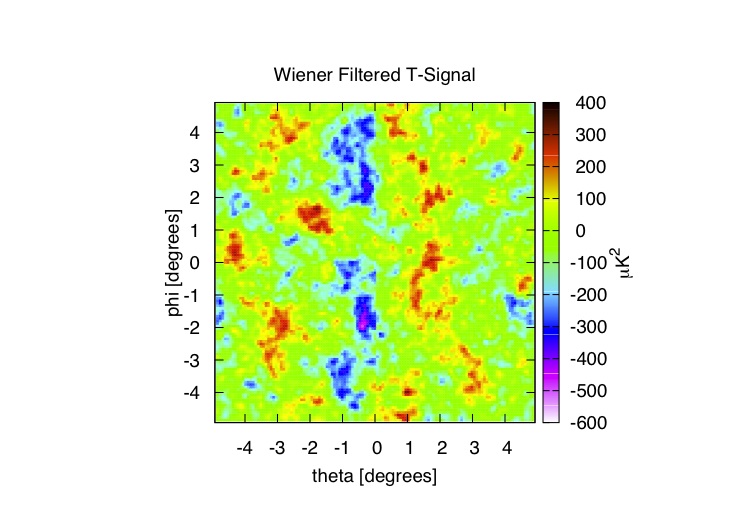} &
     \includegraphics[trim = 2cm 1cm 2cm 1cm, clip=true, width=5.5cm]{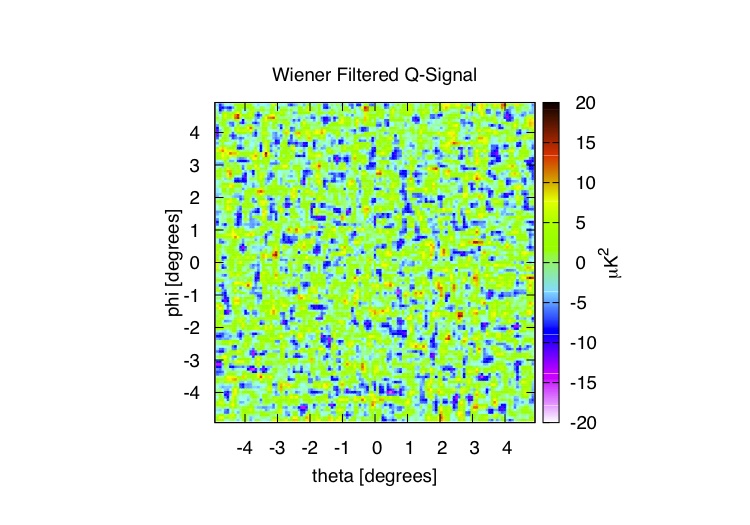} &
     \includegraphics[trim = 2cm 1cm 2cm 1cm, clip=true, width=5.5cm]{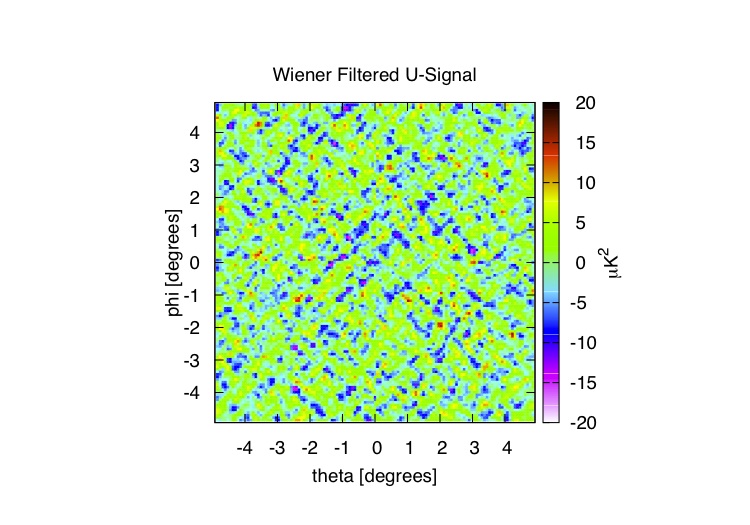} \\
     \mbox{(a)Temperature} & \mbox{(b) Stokes $Q$} & \mbox{(c) Stokes $U$}
     \end{array}$
       \caption{ Wiener filtered maps. a) Temperature, b) Stokes $Q$ and c) Stokes $U$ components of the solution of Eq.  7; $\left<F^{-1}\mathbf{R} \vec{x}\right>$, transformed into Stokes variables $T$, $Q$ and $U$ and averaged over all iterations. The Wiener filtered maps provide the information content of the data.}
  \end{center}
\end{figure*}

\begin{figure*} \label{fig:maps}
  \begin{center}$
    \leavevmode
    \begin{array}{ccc}
     \includegraphics[trim = 2cm 1cm 2cm 1cm, clip=true, width=5.5cm]{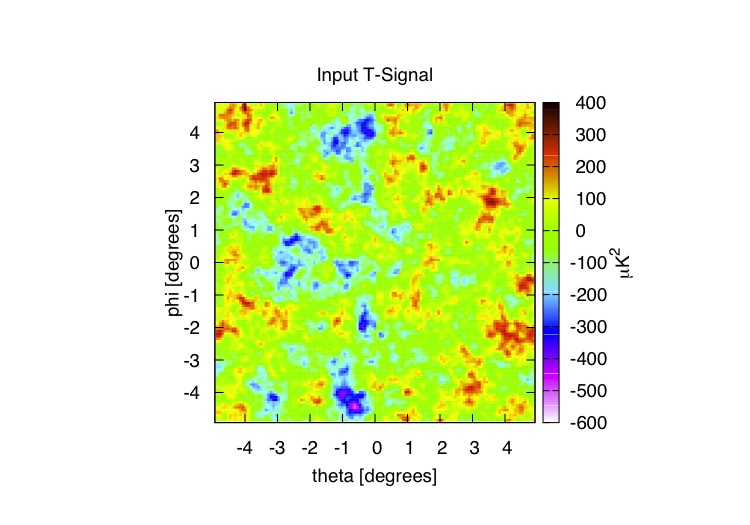} &
     \includegraphics[trim = 2cm 1cm 2cm 1cm, clip=true, width=5.5cm]{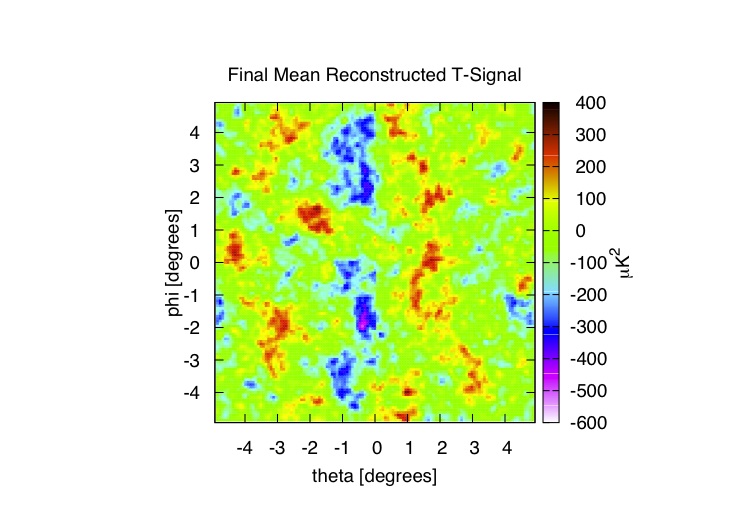} &
     \includegraphics[trim = 2cm 1cm 2cm 1cm, clip=true, width=5.5cm]{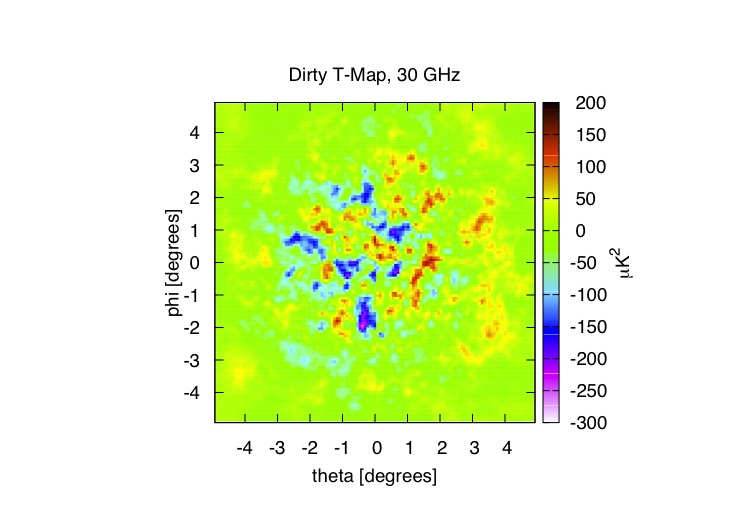} \\
     \includegraphics[trim = 2cm 1cm 2cm 1cm, clip=true, width=5.5cm]{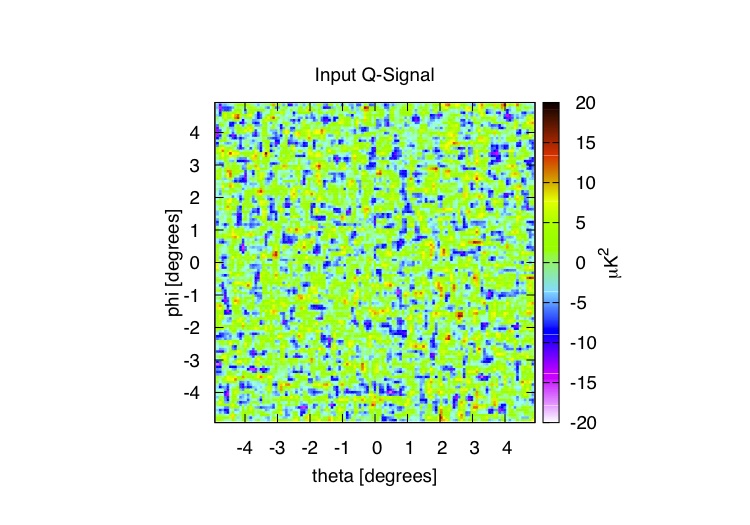} &
     \includegraphics[trim = 2cm 1cm 2cm 1cm, clip=true, width=5.5cm]{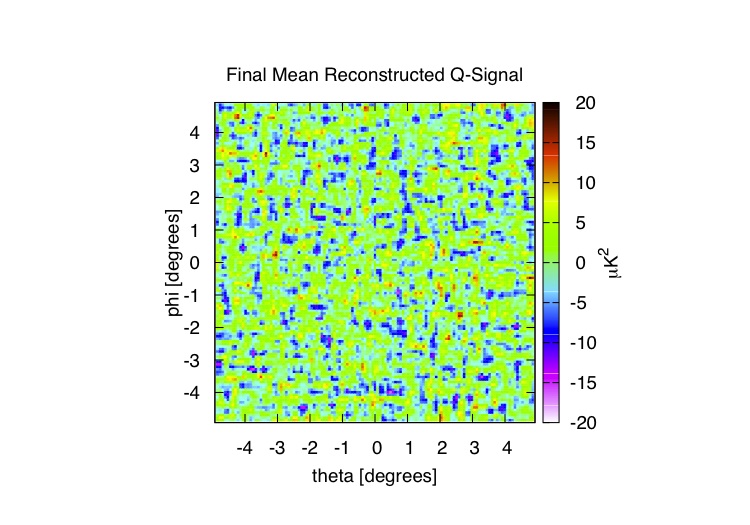} &
     \includegraphics[trim = 2cm 1cm 2cm 1cm, clip=true, width=5.5cm]{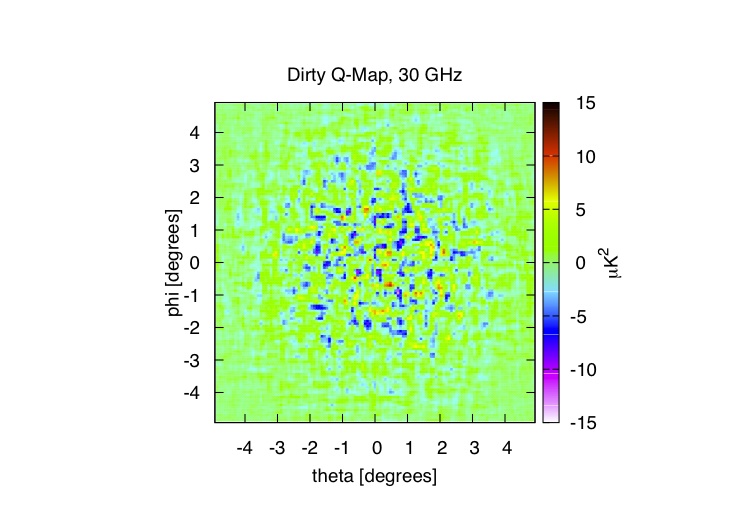} \\
     \includegraphics[trim = 2cm 1cm 2cm 1cm, clip=true, width=5.5cm]{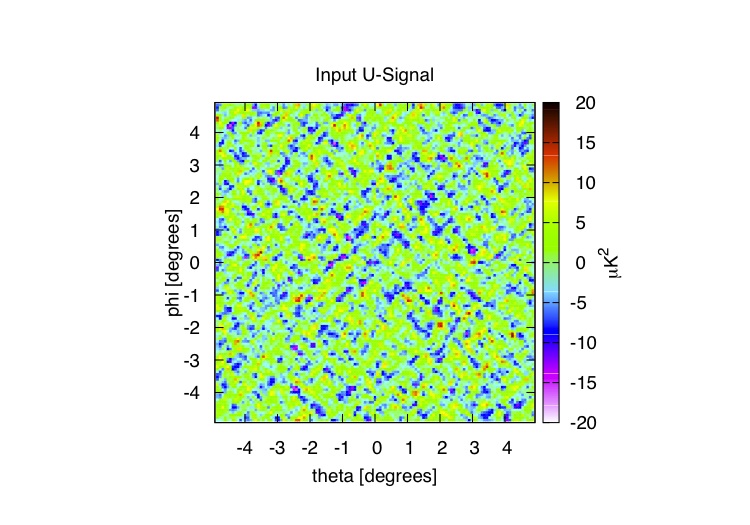} &
     \includegraphics[trim = 2cm 1cm 2cm 1cm, clip=true, width=5.5cm]{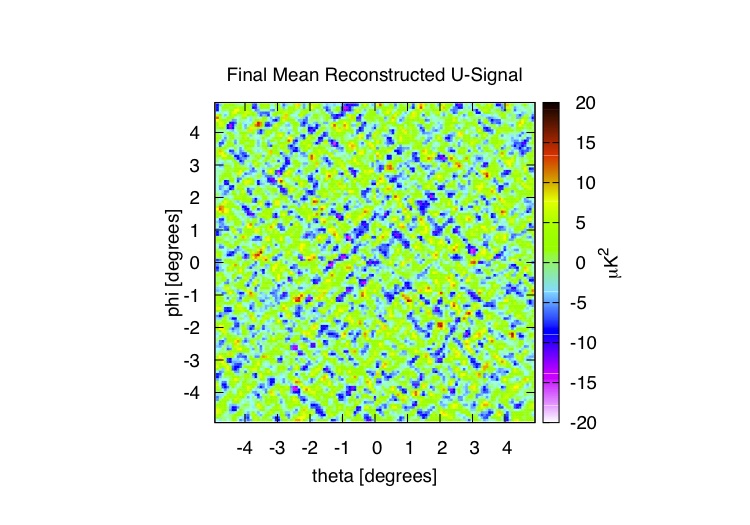} &
     \includegraphics[trim = 2cm 1cm 2cm 1cm, clip=true, width=5.5cm]{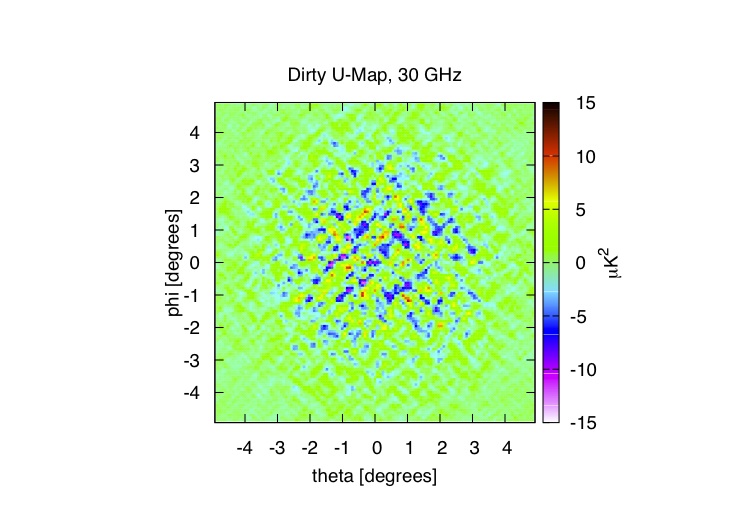} \\
     \mbox{(a) Signal Realization} & \mbox{(c) Final Mean Posterior Map} & \mbox{(c) Dirty Map} 
     \end{array}$
       \caption{Signal maps. a) The signal realization, which is constructed from the input power spectra shown in blue in Figure 4, is used as the input map for the interferometer simulation.  b) The final mean posterior map is the sum of solutions of Eq. 7 and Eq. 8; $\left< F^{-1}\mathbf{R} (\vec{x} + \vec{y}) \right>$, transformed into Stokes variables $T$, $Q$ and $U$ and averaged over all iterations. It provides the reconstruction of the noiseless input signal by the Gibbs sampler within the area of the primary beam. c) The dirty map is simply the inverse Fourier transform of the data. The three rows show, from top to bottom, temperature, Stokes $Q$ and Stokes $U$ parameters.}
  \end{center}
\end{figure*}

\section{Conclusion}

In this work the extension of Gibbs sampling to interferometric observations of polarized signals has been successfully demonstrated. An example of signal reconstruction and inference of CMB power spectra from a moderately large ($n_p = 128^2$) mock data set has been provided. The validity of our technique in dealing with realistic interferometric data, including an incomplete $uv$-plane coverage, finite beam size and baseline-dependent noise, has been shown.

A polarization signal cannot always be uniquely decomposed into E and B parts  on a cut sky. The non-uniqueness of the decomposition causes leakage from the E-mode into the much weaker B-mode power spectrum. Our input signal maps were generated on a flat sky patch with periodic boundary conditions. Because the $E$-$B$ decomposition is unique in this domain, the so-called $E$-$B$ coupling problem did not arise in the recovery of the signal realization. Since the Gibbs sampler recovers the power spectra of the signal realization, we have a good agreement between the ``input" and the ``mean posterior" spectra in Figure 4. If the signals had been produced as patches cut out from full-sky maps, then the ``realization'' and the ``mean posterior" power spectra in Figure 4 would have been severely contaminated by $E$-$B$ coupling, which can be easily resolved in the flat-sky approximation (Bunn 2002).

Our Gibbs sampling approach is also applicable to more realistic cases of interferometic polarimetry simulations, such as close-packed arrays with systematic errors. Such simulations with varying systematics can provide an idea about the limitations of interferometers for upcoming missions, such as the QUBIC experiment, which aims to detect $B$-mode polarization anisotropies of the CMB signal. 

\section*{Acknowledgments}
Computing resources were provided by the University of Richmond under NSF Grant 0922748. Our implementation of the Gibbs sampling algorithm uses the open-source PETSc library (Balay et al. 1997, 2010, 2011) and FFTW (Frigo \& Johnson 2005). G. S. Tucker and A. Karakci acknowledge support from NSF Grant AST-0908844. P. M. Sutter and B. D. Wandelt acknowledge support from NSF Grant AST-0908902. B. D. Wandelt acknowledges funding from an ANR Chaire d'Excellence, the UPMC Chaire Internationale in Theoretical Cosmology, and NSF grants AST-0908902 and AST-0708849. L. Zhang and P. Timbie acknowledge support from NSF Grant AST-0908900. E. F. Bunn acknowledges support from NSF Grant AST-0908900. We are grateful for the generous hospitality of The Ohio State University's Center for Cosmology and Astro-Particle Physics, which hosted a workshop during which some of these results were obtained.

\bibliography{polarGibbs.bbl}

\begin{thebibliography}{404}
\expandafter\ifx\csname natexlab\endcsname\relax\def\natexlab#1{#1}\fi

\bibitem[{{Balay} et~al.(1997)}]{balay97}
Balay, S., Gropp, W. D., McInnes, L. C., \& Smith, B. F. 1997,
in \emph{Modern Software Tools in Scientific Computing}, ed. E. Arge,
- Revision 3.1, Argonne National Laboratory A. M. Bruaset, \&
H. P. Langtangen, (Birkhauser Press) 163Ð 202.

\bibitem[{Balay {et~al.}(2010)}]{balay10}
Balay, S., Gropp, W. D., McInnes, L. C., \& Smith, B. F.  2010,
\emph{PETSc Users Manual} - Revision 3.1, Argonne National Laboratory
- http://www.mcs.anl.gov/petsc/

\bibitem[{Batiselli {et~al.}(2010)}]{qubic}
Battistelli, E., Baœ, A., Bennett, D., BergŽ, L., Bernard, J.-Ph., de Bernardis, P., Bordier, G., Bounab, A., BrŽelle, ƒ., Bunn, E. F., Calvo, M., Charlassier, R., Collin, S., Coppolecchia, A., Cruciani, A., Curran, G., de Petris, M., Dumoulin, L., Gault, A., Gervasi, M., Ghribi, A., Giard, M., Giordano, C., Giraud-HŽraud, Y., Gradziel, M., Guglielmi, L., Hamilton, J.-Ch., Haynes, V., Kaplan, J., Korotkov, A., LandŽ, J., Maffei, B., Maiello, M., Malu, S., Marnieros, S., Martino, J., Masi, S., Murphy, A., Nati, F., O'Sullivan, C., Pajot, F., Passerini, A., Peterzen, S., Piacentini, F., Piat, M., Piccirillo, L., Pisano, G., Polenta, G., Prle, D., Romano, D., Rosset, C., Salatino, M., Schillaci, A., Sironi, G., Sordini, R., Spinelli, S., Tartari, A., Timbie, P., Tucker, G., Vibert, L., Voisin, F., Watson, R. A., and Zannoni, M. 2010, Astroparticle Physics, Volume 34, Issue 9, 705-716.

\bibitem[Bunn, 2002]{bunn02}
Bunn, E. F. 2002, Phys. Rev. D 65, 043003

\bibitem[Bunn, 2003]{bunn03}
Bunn, E. F. 2003, New Astronomy Review 47 987-994

\bibitem[Bunn, 2007]{bunn07}
Bunn, E. F. 2007, Phys. Rev. D 75 083517

\bibitem[{Bunn {et~al.}(1994)}]{bunn94}
Bunn, E. F., Fisher, K. B., Hoffman, Y., Lahav, O., Silk, J, \& Zaroubi, S. 1994, ApJ, 432, L75

\bibitem[{Dickson {et~al.}(2004)}]{dick}
Dickinson, C. et al. 2004, MNRAS, 353, 732 Ñ. 2009, ApJ, 705, 1607

\bibitem[{Elsner \& Wandelt(2012a)}]{elwan}
Elsner, F. \& Wandelt, B. D. 2012a, A\&A, 542, A60

\bibitem[{Elsner \& Wandelt(2012b)}]{wanel}
Elsner, F. \& Wandelt, B. D. 2012b, astro-ph/1211.0585

\bibitem[{Eriksen {et~al.}(2007)}]{erik} Eriksen, H. K., Huey, G., Banday, A. J., Gorski, K. M., Jewell,
J. B., OÕDwyer, I. J., \& Wandelt, B. D. 2007, ApJ, 665, L1

\bibitem[{Frigo \& Johnson(2005)}]{frig} Frigo, M. \& Johnson, S. 2005, Proceedings of the IEEE, 93, 216

\bibitem[{Gelman {et~al.}(2004)}]{getal}
Gelman, A., Carlin, J. B., Stern, H. S., \& Rubin, D. 2004,
\emph{Bayesian Data Analysis}, (2nd ed.) Boca Raton, FL: Chapman
and Hall/CRC

\bibitem[{Gelman \& Rubin(1992)}]{gelman}
Gelman, A. \& Rubin, D. 1992, Statistical Science, 7, 457

\bibitem[{Grainge {et~al.}(2003)}]{grain}
Grainge, K. et al. 2003, MNRAS, 341, L23

\bibitem[{Hoffman \& Ribak(1991)}]{hoff}
Hoffman, Y. \& Ribak, E. 1991, Ap. J. Lett. 380, L5

\bibitem[{Hu {et~al.}(2003)}]{huza}
Hu, W., Hedman, M. M. and Zaldarriaga, M. 2003, Phys. Rev. D
67 043004

\bibitem[{Hu \& White(1997)}]{huh}
Hu, W. \& White, M. 1997, New Astronomy 2:323-344

\bibitem[{Jewell {et~al.}(2004)}]{jew}
Jewell, J., Levin, S., \& Anderson, C. H. 2004, ApJ, 609, 1

\bibitem[{Kamionkowski {et~al.}(1997)}]{kam}
Kamionkowski, M, Kososwsky, A. \& Stebbins, A. 1997, Phys.
Rev. D 55 7368-7388

\bibitem[{Komatsu {et~al.}(2011)}]{wmapk}
Komatsu, E. et al. 2011, ApJS 192, 18

\bibitem[{Kovac {et~al.}(2002)}]{dasi}
Kovac, J. M., Leitch, E. M., Pryke, C., Carlstrom, J. E.,
Halverson, N. W., \& Holzapfel, W. L. 2002, Nature, 420, 772

\bibitem[{Larson {et~al.}(2007)}]{lar}
Larson, D. L., Eriksen, H. K., Wandelt, B. D., Gorski, K. M.,
Huey, G., Jewell, J. B., and O'Dwyer, I. J. 2007, ApJ, 656, 653

\bibitem[{Larson {et~al.}(2011)}]{wmapl}
Larson, D. et al. 2011, ApJS 192, 16

\bibitem[{Lewis {et~al.}(2000)}]{camb}
Lewis, A., Challinor, A., \& Lasenby, A. 2000, ApJ 538, 473

\bibitem[{Lewis {et~al.}(2002)}]{lewis}
Lewis, A., Challinor, A. \& Turok, N. 2002, Phys. Rev. D 65
023505

\bibitem[{O`Dwyer {et~al.}(200d)}]{odw} 
OÕDwyer, I. J., Eriksen, Wandelt, B. D., Jewell, J. B., 
Larson, D. L., Gorski, K. M., G., Banday, A. J., Lewin, S. \&  Lilje. P. B. 2004, ApJ, 617, L99

\bibitem[{Park {et~al.}(2003)}]{parkal}
Park, C. G., Ng, K. W., Park, C., Liu, C. G., \& Umetsu, K. 2003,
ApJ 589 67-81

\bibitem[{Park \& Ng(2004)}]{parng}
Park, C. G., Ng, K. W. 2004, ApJ 609 15-21

\bibitem[{Parson {et~al.}(2003)}]{cbi}
Pearson, T. J., et al. 2003, ApJ 591, 556-574

\bibitem[{Press {et~al.}(1986)}]{press}
Press, W. H., Flannery, Brian P., \& Teukolsky, Saul A. 1986, \emph{Numerical Recipes: The Art of Scientific Computing},
Cambridge University Press

\bibitem[{Sutter {et~al.}(2012)}]{sut}
Sutter, P. M., Wandelt, B. D., Malu, S. S., 2012 ApJS 202 9,
astro-ph/1109.4640

\bibitem[{Wandelt {et~al.}(2004)}]{wand}
Wandelt, B. D., Larson, D. L., and Lakshminarayanan, A. 2004,
Phys. Rev. D, 70, 083511

\bibitem[{White {et~al.}(1999)}]{white}
White, M., Carlstrom, J. E., Dragovan, M., \& Holzapfel, W. L. 1999, ApJ, 514, 12 

\bibitem[{Zaldarriaga \& Seljak(1997)}]{zald}
Zaldarriaga, M. \& Seljak, U. 1997, Phys. Rev. D 55 1830-1840

\bibitem[{Zaldarriaga \& Seljak(1998)}]{selj}
Zaldarriaga, M. \& Seljak, U. 1998, Phys. Rev. D 58 023003
  
\end{thebibliography}
\bibliographystyle{apj}
\nocite{*}

\end{document}